\documentclass[useAMS,usegraphicx,usenatbib,a4paper]{mn2e}
\usepackage{color}
\usepackage{wasysym}
\usepackage{url}
\usepackage{multirow}
\usepackage[tight]{subfigure}
\usepackage{graphicx}
\usepackage{verbatim}
\usepackage{times}
\usepackage[usenames,dvipsnames,table]{xcolor}

\newcommand{\msun}{\ensuremath{\mathrm{M}_\odot}}

\def\kms{km\,s$^{-1}$}
\def\HII{H\,{\sc ii}}

\def\CII{C\,{\sc ii}}
\def\NII{N\,{\sc ii}}
\def\OI{O\,{\sc i}}

\def\OII{O\,{\sc ii}}

\def\OIII{O\,{\sc iii}}

\def\SII{S\,{\sc ii}}

\def\SiII{Si\,{\sc ii}}

\def\CaII{Ca\,{\sc ii}}
\def\FeI{Fe\,{\sc i}}
\def\FeII{Fe\,{\sc ii}}
\def\FeIII{Fe\,{\sc iii}}
\def\FeIV{Fe\,{\sc iv}}

\def\NiII{Ni\,{\sc ii}}

\def\CaII{Ca\,{\sc ii}}

\def\Nifs{$^{56}$Ni}
\def\Cofs{$^{56}$Co}

\def\dm15{$\Delta m_{15}(B)$}
\def\MCh{M$_\mathrm{Ch}$}
\def\lesssim{\mathrel{\hbox{\rlap{\hbox{\lower4pt\hbox{$\sim$}}}\hbox{$<$}}}}
\let\la=\lesssim
\def\gtrsim{\mathrel{\hbox{\rlap{\hbox{\lower4pt\hbox{$\sim$}}}\hbox{$>$}}}}
\let\ga=\gtrsim

\def\aj{AJ}%
\def\apj{ApJ}%
\def\apjl{ApJ}%
\def\aap{A\&A}%
\def\aaps{A\&AS}%
\def\mnras{MNRAS}%
\def\pasp{PASP}%
\def\pasj{PASJ}%
\def\nat{Nature}%
\def\procspie{Proc.~SPIE}%
\title[SN~2012dn]{SN\,2012dn from early to late times: 09dc-like supernovae reassessed\thanks{Based on observations collected with ESO telescopes under programme IDs 091.D-0600 and 290.D-5035.}}
\author[Taubenberger et al.]{S.~Taubenberger$^{1,2}$, %\thanks{E-mail: tauben@mpa-garching.mpg.de}, 
A.~Floers$^{1,2}$, C.~Vogl$^{2}$, M.~Kromer$^{3,4}$, J.~Spyromilio$^{1}$, G.~Aldering$^{5}$,
\newauthor P.~Antilogus$^{6}$, S.~Bailey$^{5}$, C.~Baltay$^{7}$, S.~Bongard$^{6}$, K.~Boone$^{5,8}$, C.~Buton$^{9}$, N.~Chotard$^{9}$, 
\newauthor Y.~Copin$^{9}$, S.~Dixon$^{8}$, D.~Fouchez$^{10}$, C.~Fransson$^{11}$, E.~Gangler$^{12}$, R.\,R.~Gupta$^{5}$, 
\newauthor S.~Hachinger$^{13}$, B.~Hayden$^{5}$, W. Hillebrandt$^{2}$, A.\,G.~Kim$^{5}$, M.~Kowalski$^{14,15}$, P.-F.~Leget$^{12}$, 
\newauthor B.~Leibundgut$^{1}$, P.\,A.~Mazzali$^{2,16}$, U.\,M.~Noebauer$^{2}$, J.~Nordin$^{14}$, R.~Pain$^{6}$, R.~Pakmor$^{2,4}$, 
\newauthor E.~Pecontal$^{17}$, R.~Pereira$^{9}$, S.~Perlmutter$^{5,8}$, K.\,A.~Ponder$^{18}$, D.~Rabinowitz$^{7}$, M.~Rigault$^{12}$, 
\newauthor D.~Rubin$^{5,19}$, K.~Runge$^{5}$, C.~Saunders$^{5,20}$, G.~Smadja$^{9}$, C.~Tao$^{10,21}$, R.\,C.~Thomas$^{22}$\\ 
$^{1}$European Southern Observatory, Karl-Schwarzschild-Str.~2, D-85748 Garching, Germany\\
$^{2}$Max-Planck-Institut f\"{u}r Astrophysik, Karl-Schwarzschild-Str.~1, D-85741 Garching, Germany\\
$^{3}$Zentrum f\"{u}r Astronomie der Universit\"{a}t Heidelberg, Institut f\"{u}r Theoretische Astrophysik, D-69120 Heidelberg, Germany\\
$^{4}$Heidelberger Institut f\"{u}r Theoretische Studien, D-69118 Heidelberg, Germany\\
$^{5}$Physics Division, Lawrence Berkeley National Laboratory, 1 Cyclotron Road, Berkeley, CA 94720, USA\\
$^{6}$Laboratoire de Physique Nucl\'eaire et de Hautes Energies, CNRS/IN2P3, Sorbonne Univ., Univ. Paris Diderot, 4 Place Jussieu, F-75252 Paris Cedex 05, France\\
$^{7}$Department of Physics, Yale University, New Haven, CT 06250-8121, USA\\
$^{8}$Department of Physics, University of California Berkeley, 366 LeConte Hall MC 7300, Berkeley, CA 94720-7300, USA\\
$^{9}$Institut de Physique Nucl\'eaire de Lyon, CNRS/IN2P3, Universit\'e de Lyon, Universit\'e de Lyon 1, F-69622 Villeurbanne Cedex, France\\
$^{10}$Centre de Physique des Particules de Marseille, CNRS/IN2P3, Aix Marseille Universit\'e, F-13288 Marseille Cedex 09, France\\
$^{11}$The Oskar Klein Centre and Department of Astronomy, Stockholm University, Albanova, SE-10691 Stockholm, Sweden\\
$^{12}$Laboratoire de Physique de Clermont, CNRS/IN2P3, Universit\'e Clermont Auvergne, F-63000 Clermont-Ferrand, France\\
$^{13}$Leibniz Supercomputing Centre (LRZ), Bavarian Academy of Sciences and Humanities, Boltzmannstr. 1, D-85748 Garching, Germany\\
$^{14}$Institut f\"{u}r Physik, Humboldt-Universit\"{a}t zu Berlin, Newtonstr. 15, D-12489 Berlin, Germany\\
$^{15}$DESY, D-15735 Zeuthen, Germany\\
$^{16}$Astrophysics Research Institute, Liverpool John Moores University, IC2, Liverpool Science Park, 146 Brownlow Hill, Liverpool L3 5RF, UK\\
$^{17}$Centre de Recherche Astronomique de Lyon, Universit\'e Lyon 1, 9 Avenue Charles Andr\'e, F-69561 Saint Genis Laval Cedex, France\\
$^{18}$Berkeley Center for Cosmological Physics, University of California Berkeley, 341 Campbell Hall, Berkeley, CA 94720, USA\\
$^{19}$Space Telescope Science Institute, 3700 San Martin Drive, Baltimore, MD 21218, USA\\
$^{20}$Institut Lagrange de Paris (ILP), Sorbonne Universit\'es, 98 bis Boulevard Arago, F-75014 Paris, France\\
$^{21}$Tsinghua Center for Astrophysics, Tsinghua University, Beijing 100084, China\\
$^{22}$Computational Research Division, Lawrence Berkeley National Laboratory, 1 Cyclotron Road MS 50B-4206, Berkeley, CA 94720, USA
}

\begin{document}

\date{Accepted 2019 June 11. Received 2019 June 10; in original form 2019 March 27}

\pagerange{\pageref{firstpage}--\pageref{lastpage}} \pubyear{2019}

\maketitle

\label{firstpage}

\begin{abstract}
As a candidate `super-Chandrasekhar' or 09dc-like Type~Ia supernova (SN~Ia), SN~2012dn shares many characteristics with other members of this remarkable class of objects but lacks their extraordinary luminosity. Here, we present and discuss the most comprehensive optical data set of this SN to date, comprised of a densely sampled series of early-time spectra obtained within the Nearby Supernova Factory project, plus photometry and spectroscopy obtained at the Very Large Telescope about 1\,yr after the explosion. The light curves, colour curves, spectral time series and ejecta velocities of SN~2012dn are compared with those of other 09dc-like and normal SNe~Ia, the overall variety within the class of 09dc-like SNe~Ia is discussed, and new criteria for 09dc-likeness are proposed. Particular attention is directed to additional insight that the late-phase data provide. The nebular spectra show forbidden lines of oxygen and calcium, elements that are usually not seen in late-time spectra of SNe~Ia, while the ionisation state of the emitting iron plasma is low, pointing to low ejecta temperatures and high densities. The optical light curves are characterised by an enhanced fading starting $\sim$60\,d after maximum and very low luminosities in the nebular phase, which is most readily explained by unusually early formation of clumpy dust in the ejecta. Taken together, these effects suggest a strongly perturbed ejecta density profile, which might lend support to the idea that 09dc-like characteristics arise from a brief episode of interaction with a hydrogen-deficient envelope during the first hours or days after the explosion.
\end{abstract}


\section{Introduction}
\label{Introduction}

When the first `super-Chandrasekhar' Type~Ia supernova (SN~Ia) was identified \citep[SN~2003fg;][]{howell2006a,branch2006b}, the outstanding characteristics were a luminosity twice as high as in ordinary SNe~Ia, and comparatively low ejecta velocities. It was readily realised that, within the paradigm of radioactivity-driven light curves, SN~2003fg could not be consistent with a total ejecta mass limited to 1.4\,\msun, the Chandrasekhar-mass (\MCh) stability limit of non-rotating white dwarfs (WDs), since the required \Nifs\ mass alone would be close to that number. This is essentially how the name `super-Chandrasekhar' SNe~Ia came into being, since the initially proposed solution to the problem was a total ejecta mass far in excess of \MCh. It was suggested that this could be achieved with thermonuclear explosions of rapidly rotating, supermassive WDs \citep{howell2006a,branch2006b}, possibly as the result of a WD merger \citep{hicken2007a}.

However, the model of exploding super-Chandrasekhar-mass WDs has been severely challenged both by numerical explosion simulations \citep{pfannes2010a,pfannes2010b,fink2018a} and the analysis of the more complete data sets available for the `super-Chandrasekhar' SNe~2006gz \citep{hicken2007a,maeda2009a}, 2007if \citep{scalzo2010a,scalzo2012a,yuan2010a} and 2009dc \citep{yamanaka2009a,silverman2011a,taubenberger2011a,taubenberger2013a,hachinger2012a}. As a consequence, other scenarios now appear to offer more promising explanations for these events \citep{taubenberger2017a}, and we argue that they should better be named in a less interpretative manner. We propose the term `09dc-like SNe~Ia', following the nomenclature typically used to denote SN subclasses based on spectroscopic and photometric similarity to a prototypical event, which SN~2009dc is often considered to be in this case. What this involves in detail will be elaborated further in Section~\ref{Defining characteristics of the class of 09dc-like SNe Ia}.

With the additional data, significant diversity has become apparent within the group of well-studied 09dc-like SNe~Ia in terms of early-time spectroscopic appearance, peak luminosity, ejecta velocities and late-time behaviour, and not all authors apply the same criteria to identify potential members of the class. For instance, \citet{scalzo2012a,scalzo2014c} studied a group of objects with strong spectroscopic similarities to SN~2007if at early phases (e.g., strong \FeIII\ absorption lines), but much less extreme photometric properties, and suggested them to result from the tamped-detonation scenario (i.e., the detonation of a WD inside an extended envelope; \citealt{khokhlov1993b,hoeflich1996a}). Most of these objects are classical 91T-like SNe. We will show in Section~\ref{Defining characteristics of the class of 09dc-like SNe Ia} that 91T-like SNe differ in several respects from 09dc-like events as defined here.

From a comparison of several 09dc-like objects,
a common theme arises: a suite of properties that many of them share, even though the agreement is often only partial. In this process the supposed hallmark features of an extreme luminosity and low ejecta velocities have lost some significance, since they are not as universal as initially thought. Instead, peculiarities such as unusually strong and persistent carbon lines, an unusual light-curve morphology and a peculiar late-time behaviour with low ejecta ionisation have become equally important connecting links. 
SN~2012dn currently marks the extreme in this respect: its luminosity is perfectly in line with that of normal SNe~Ia, and its ejecta velocities are not particularly low either. Still, it is clearly related to other 09dc-like events on the basis of the other properties mentioned above. 

SN~2012dn was discovered on \textsc{UT} 2012 July 08.52 at $\alpha = 20^\mathrm{h}23^\mathrm{m}36\fs26$ and $\delta = -28\degr16\arcmin43\farcs4$ (J2000) in the SA(s)cd spiral galaxy ESO 462-016 \citep{bock2012a}. Based on spectra taken on \textsc{UT} 2012 July 10.2 and 11.5 at the Gemini South and UH 88-inch telescopes, respectively, SN~2012dn was classified as an SN~Ia exhibiting strong similarities to SN~2006gz \citep{hicken2007a} a week to ten days before maximum light \citep{parrent2012a,copin2012a}. Since then, a number of studies on various aspects of SN~2012dn have been published. \citet{chakradhari2014a} and \citet{parrent2016a} focused on the early-time photometric and spectroscopic properties at optical wavelengths, emphasising the similarities of SN~2012dn with other 09dc-like events despite its comparatively low luminosity. Spectral modelling with SYNAPPS \citep{thomas2011a} was carried out by \citet{parrent2016a}. \citet{brown2014a} remarked on the unusually high UV flux of SN~2012dn and other 09dc-like SNe~Ia at early phases. \citet{yamanaka2016a} found a near-IR (NIR) excess in SN~2012dn starting about one month after maximum, interpreted as a NIR echo on pre-existing circumstellar dust \citep{yamanaka2016a,nagao2017a,nagao2018a}.

Here, we present and analyse the extensive set of optical spectrophotometric observations collected by the Nearby Supernova Factory (SNfactory; \citealt{aldering2002a}), covering the epochs from 13\,d before to 102\,d after the time of $B$-band maximum. This data set is complemented by additional optical photometry and spectroscopy obtained at the VLT $\sim$1\,yr after the explosion, allowing us for the first time to assess the entire evolution of SN~2012dn all the way into the nebular phase.
The paper is organised as follows. In Section~\ref{Observations of SN 2012dn and data reduction} the new observations of SN~2012dn are presented and the methods of data reduction are detailed. In Section~\ref{Analysis of early-time data} the early-time photometric and spectroscopic data are discussed and compared to those of other 09dc-like events as well as normal SNe~Ia. Section~\ref{Analysis of late-time data} is devoted to late-time data taken $\sim$1\,yr after the explosion, highlighting the new insights that can be derived from these observations. A general discussion about the possible origin of an accelerated late-time light-curve decline and the observed NIR excess, their implications for possible explosion models, and a systematic group definition of 09dc-like SNe~Ia follow in Section~\ref{Discussion}. The paper concludes with a short summary of the main results in Section~\ref{Conclusions}.

\section{Observations of SN 2012dn and data reduction}
\label{Observations of SN 2012dn and data reduction}

\subsection{SNIFS observations}
\label{SNIFS observations}

Immediately after the discovery of SN~2012dn, SNfactory initiated an intense follow-up campaign with roughly a 3-day cadence over 120\,d at the UH 88-inch Telescope on Mauna Kea, Hawaii, equipped with the SNIFS spectrograph \citep{lantz2004a}. SNIFS is an integral-field spectrograph \`a la Tiger \citep{bacon1995a}, specifically designed for accurate spectrophotometry of point sources superimposed on a structured background. A field of view of $6.4 \times 6.4$\, arcsec$^2$ is sampled by an array of $15 \times 15$ microlenses, resulting in an image scale of 0.43\,arcsec per spatial element. The light from each spatial element is split into two beams by a dichroic at 5100\,\AA. Each beam is directed onto a grating and recorded by a CCD optimised for the respective wavelength range.

The SNfactory data-reduction pipeline has been discussed by \citet{aldering2006a} and \citet{scalzo2010a}. CCD frames are wavelength-calibrated with the help of arc-lamp exposures taken immediately after every science exposure with the same telescope pointing to establish the microlens point-spread function (PSF) on the detector and to eliminate inaccuracies due to flexure in the telescope or instrument. The data are then mapped into (x,y,$\lambda$) data cubes. In order to cleanly subtract any non-uniform background emission underlying the SN, a deep spectrophotometric exposure at the SN location is taken once the SN itself has faded below detectability. In the case of SN~2012dn, the template was acquired in 2014, approximately two years after the SN explosion. This template exposure is spatially registered to the earlier SN exposures and subtracted off at the data-cube level \citep{bongard2011a}. Spectra are extracted from the host-subtracted data cubes using a spatial PSF, subtracting residual sky emission as a spatially uniform background.

Flux calibration and a correction of the atmospheric extinction \citep{buton2013a} are accomplished through the observation of a number of spectrophotometric standard stars on every night. To ensure an accurate absolute flux calibration even in non-photometric nights, a field adjacent to the SN is always imaged simultaneously through the SNIFS multi-filters photometric channel during the spectral exposures. By comparing the stellar fluxes measured during non-photometric nights with those calibrated under photometric conditions, a correction for any grey differential atmospheric attenuation between the observations of the SN and the standard stars is obtained. The per-object reproducibility thus achieved is better than $\sim$2 per cent throughout the observed wavelength window of 3300 to 9700\,\AA.

\begin{table*}
\caption{Restframe Bessell $U\!BV\!RI$ magnitudes of SN~2012dn. Phases are given in restframe days with respect to the $B$-band peak (MJD = $56132.6\pm0.2$). } 
\begin{footnotesize}
\begin{center}
\begin{tabular}{crcccccl}
\hline
MJD & Phase & $U$ & $B$ & $V$ & $R$ & $I$ & Source\\
\hline
\vspace{0.04cm}
56119.48  & $-13.0$ & $15.02 \pm 0.15$ & $15.83 \pm 0.07$ & $15.74 \pm 0.06$ & $15.57 \pm 0.04$ & $15.48 \pm 0.04$ &  SNIFS \\
\vspace{0.04cm}
56121.44  & $-11.1$ & $14.44 \pm 0.14$ & $15.23 \pm 0.06$ & $15.16 \pm 0.05$ & $15.01 \pm 0.04$ & $14.95 \pm 0.04$ &  SNIFS \\
\vspace{0.04cm}
56123.48  &  $-9.0$ & $14.15 \pm 0.14$ & $14.91 \pm 0.06$ & $14.85 \pm 0.05$ & $14.73 \pm 0.04$ & $14.72 \pm 0.04$ &  SNIFS \\
\vspace{0.04cm}
56126.42  &  $-6.1$ & $13.87 \pm 0.13$ & $14.59 \pm 0.06$ & $14.57 \pm 0.05$ & $14.47 \pm 0.04$ & $14.52 \pm 0.04$ &  SNIFS \\
\vspace{0.04cm}
56128.42  &  $-4.1$ & $13.79 \pm 0.13$ & $14.45 \pm 0.06$ & $14.41 \pm 0.05$ & $14.31 \pm 0.04$ & $14.38 \pm 0.04$ &  SNIFS \\
\vspace{0.04cm}
56131.43  &  $-1.2$ & $13.84 \pm 0.12$ & $14.40 \pm 0.06$ & $14.34 \pm 0.05$ & $14.23 \pm 0.04$ & $14.31 \pm 0.04$ &  SNIFS \\
\vspace{0.04cm}
56133.44  &    0.8  & $13.93 \pm 0.12$ & $14.39 \pm 0.06$ & $14.31 \pm 0.05$ & $14.20 \pm 0.04$ & $14.28 \pm 0.04$ &  SNIFS \\
\vspace{0.04cm}
56136.43  &    3.8  & $14.06 \pm 0.11$ & $14.43 \pm 0.06$ & $14.28 \pm 0.05$ & $14.17 \pm 0.04$ & $14.25 \pm 0.04$ &  SNIFS \\
\vspace{0.04cm}
56141.40  &    8.7  & $14.56 \pm 0.10$ & $14.74 \pm 0.06$ & $14.42 \pm 0.05$ & $14.30 \pm 0.04$ & $14.33 \pm 0.04$ &  SNIFS \\
\vspace{0.04cm}
56143.41  &   10.7  & $14.74 \pm 0.10$ & $14.88 \pm 0.06$ & $14.44 \pm 0.05$ & $14.32 \pm 0.03$ & $14.33 \pm 0.04$ &  SNIFS \\
\vspace{0.04cm}
56146.39  &   13.6  & $15.06 \pm 0.10$ & $15.16 \pm 0.06$ & $14.55 \pm 0.05$ & $14.41 \pm 0.03$ & $14.35 \pm 0.04$ &  SNIFS \\
\vspace{0.04cm}
56148.43  &   15.7  & $15.36 \pm 0.10$ & $15.38 \pm 0.06$ & $14.65 \pm 0.05$ & $14.47 \pm 0.03$ & $14.37 \pm 0.04$ &  SNIFS \\
\vspace{0.04cm}
56151.40  &   18.6  & $15.78 \pm 0.10$ & $15.73 \pm 0.06$ & $14.82 \pm 0.05$ & $14.55 \pm 0.03$ & $14.36 \pm 0.04$ &  SNIFS \\
\vspace{0.04cm}
56156.39  &   23.5  & $16.39 \pm 0.10$ & $16.26 \pm 0.06$ & $15.11 \pm 0.05$ & $14.69 \pm 0.03$ & $14.38 \pm 0.04$ &  SNIFS \\
\vspace{0.04cm}
56158.37  &   25.5  & $16.51 \pm 0.10$ & $16.44 \pm 0.06$ & $15.22 \pm 0.05$ & $14.77 \pm 0.03$ & $14.42 \pm 0.04$ &  SNIFS \\
\vspace{0.04cm}
56161.36  &   28.5  & $16.81 \pm 0.10$ & $16.67 \pm 0.06$ & $15.41 \pm 0.05$ & $14.90 \pm 0.03$ & $14.51 \pm 0.04$ &  SNIFS \\
\vspace{0.04cm}
56163.36  &   30.4  & $16.92 \pm 0.10$ & $16.79 \pm 0.06$ & $15.51 \pm 0.05$ & $14.99 \pm 0.03$ & $14.58 \pm 0.04$ &  SNIFS \\
\vspace{0.04cm}
56166.34  &   33.4  & $17.11 \pm 0.10$ & $16.94 \pm 0.06$ & $15.63 \pm 0.05$ & $15.10 \pm 0.03$ & $14.66 \pm 0.04$ &  SNIFS \\
\vspace{0.04cm}
56171.36  &   38.4  & $17.36 \pm 0.11$ & $17.13 \pm 0.06$ & $15.80 \pm 0.05$ & $15.27 \pm 0.03$ & $14.82 \pm 0.04$ &  SNIFS \\
\vspace{0.04cm}
56173.34  &   40.3  & $17.44 \pm 0.09$ & $17.22 \pm 0.06$ & $15.94 \pm 0.05$ & $15.44 \pm 0.03$ & $14.97 \pm 0.04$ &  SNIFS \\
\vspace{0.04cm}
56176.32  &   43.3  & $17.46 \pm 0.10$ & $17.29 \pm 0.06$ & $16.02 \pm 0.05$ & $15.52 \pm 0.03$ & $15.06 \pm 0.04$ &  SNIFS \\
\vspace{0.04cm}
56178.31  &   45.2  & $17.58 \pm 0.10$ & $17.40 \pm 0.06$ & $16.12 \pm 0.05$ & $15.63 \pm 0.03$ & $15.16 \pm 0.04$ &  SNIFS \\
\vspace{0.04cm}
56181.32  &   48.2  & $17.65 \pm 0.09$ & $17.45 \pm 0.05$ & $16.19 \pm 0.05$ & $15.72 \pm 0.03$ & $15.26 \pm 0.04$ &  SNIFS \\
\vspace{0.04cm}
56183.33  &   50.2  & $17.74 \pm 0.10$ & $17.50 \pm 0.05$ & $16.24 \pm 0.05$ & $15.78 \pm 0.03$ & $15.32 \pm 0.04$ &  SNIFS \\
\vspace{0.04cm}
56186.32  &   53.2  & $17.82 \pm 0.09$ & $17.56 \pm 0.06$ & $16.31 \pm 0.05$ & $15.87 \pm 0.03$ & $15.42 \pm 0.04$ &  SNIFS \\
\vspace{0.04cm}
56188.33  &   55.2  & $17.97 \pm 0.09$ & $17.66 \pm 0.06$ & $16.40 \pm 0.05$ & $15.96 \pm 0.03$ & $15.51 \pm 0.04$ &  SNIFS \\
\vspace{0.04cm}
56191.33  &   58.1  & $18.10 \pm 0.09$ & $17.77 \pm 0.06$ & $16.53 \pm 0.05$ & $16.11 \pm 0.03$ & $15.67 \pm 0.04$ &  SNIFS \\
\vspace{0.04cm}
56193.32  &   60.1  & $18.07 \pm 0.09$ & $17.80 \pm 0.06$ & $16.61 \pm 0.05$ & $16.19 \pm 0.03$ & $15.76 \pm 0.04$ &  SNIFS \\
\vspace{0.04cm}
56196.28  &   63.0  & $18.36 \pm 0.08$ & $17.96 \pm 0.06$ & $16.72 \pm 0.05$ & $16.29 \pm 0.03$ & $15.86 \pm 0.04$ &  SNIFS \\
\vspace{0.04cm}
56198.31  &   65.0  &                  &                  & $16.83 \pm 0.05$ & $16.40 \pm 0.03$ & $15.97 \pm 0.04$ &  SNIFS \\
\vspace{0.04cm}
56206.27  &   72.9  & $18.64 \pm 0.09$ & $18.19 \pm 0.06$ & $17.11 \pm 0.05$ & $16.77 \pm 0.03$ & $16.38 \pm 0.04$ &  SNIFS \\
\vspace{0.04cm}
56208.24  &   74.9  & $18.71 \pm 0.09$ & $18.27 \pm 0.06$ & $17.18 \pm 0.05$ & $16.85 \pm 0.03$ & $16.47 \pm 0.04$ &  SNIFS \\
\vspace{0.04cm}
56211.24  &   77.8  & $18.79 \pm 0.10$ & $18.34 \pm 0.06$ & $17.31 \pm 0.05$ & $16.99 \pm 0.04$ & $16.63 \pm 0.04$ &  SNIFS \\
\vspace{0.04cm}
56213.24  &   79.8  & $18.94 \pm 0.09$ & $18.42 \pm 0.06$ & $17.39 \pm 0.05$ & $17.10 \pm 0.04$ & $16.73 \pm 0.04$ &  SNIFS \\
\vspace{0.04cm}
56218.23  &   84.8  & $19.16 \pm 0.09$ & $18.60 \pm 0.07$ & $17.61 \pm 0.05$ & $17.33 \pm 0.04$ & $16.98 \pm 0.04$ &  SNIFS \\
\vspace{0.04cm}
56228.23  &   94.7  &                  &                  & $17.97 \pm 0.06$ & $17.78 \pm 0.04$ & $17.47 \pm 0.04$ &  SNIFS \\
\vspace{0.04cm}
56236.24  &  102.6  & $19.68 \pm 0.12$ & $19.06 \pm 0.07$ & $18.24 \pm 0.07$ & $18.08 \pm 0.04$ & $17.77 \pm 0.05$ &  SNIFS \\
\vspace{0.04cm}
56423.31  &  287.8  &                  & $22.64 \pm 0.50$ & $22.61 \pm 0.24$ & $22.69 \pm 0.22$ & $21.78 \pm 0.27$ &  FORS  \\
\hline\\[-0.7ex]
\end{tabular}
\end{center}
\end{footnotesize}
\label{mags}
\end{table*}

\begin{table}
\caption{Observer-frame 2MASS $JH$ magnitudes of SN~2012dn. Phases are given in restframe days with respect to the $B$-band peak (MJD = $56132.6\pm0.2$).} 
\begin{footnotesize}
\begin{center}
\begin{tabular}{crcccccl}
\hline
MJD & Phase & $J$ & $H$ & Source\\
\hline
56507.3  & 370.9 & $22.32 \pm 0.32$ & $21.60 \pm 0.41$ &  ISAAC \\
\hline\\[-0.7ex]
\end{tabular}
\end{center}
\end{footnotesize}
\label{IRmags}
\end{table}

Finally, magnitudes in any photometric system can be calculated by integration of the flux-calibrated spectra over the respective passbands. Photometric errors can be propagated from the variance spectra. For SN~2012dn we have chosen to report SN restframe magnitudes ($z = 0.010187$) in the standard \citet{bessell2012a} photometric system to facilitate the comparison with other SNe and with the SN~2012dn photometry already published by \citet{chakradhari2014a}, \citet{brown2014a}, \citet{parrent2016a} and \citet{yamanaka2016a}. 
The blue end of the Bessell $U$-band filter is not fully covered by SNIFS spectra. Before measuring synthetic $U$-band magnitudes, we therefore had to extrapolate our spectra by 250\,\AA\ to the blue of their minimum wavelength $\lambda_\mathrm{min}$. This was done assuming constant flux, corresponding to the average flux level in the interval [$\lambda_\mathrm{min}$; $\lambda_\mathrm{min}$\,+\,200\,\AA]. The difference to an extrapolation assuming zero flux is between 0.02 and 0.07\,mag, depending on the epoch (and hence colour) of the SN. As a conservative estimate, we added this difference linearly to the $U$-band error based on the variance spectra.
The Bessell $U\!BV\!RI$ magnitudes from SNIFS spectrophotometry, along with their uncertainties, are reported in Table~\ref{mags}. The spectra are available on the SNfactory website, http://snfactory.lbl.gov/snf/data/index.html.

\subsection{VLT observations}
\label{VLT observations}

\begin{figure}
  \centering 
  \includegraphics{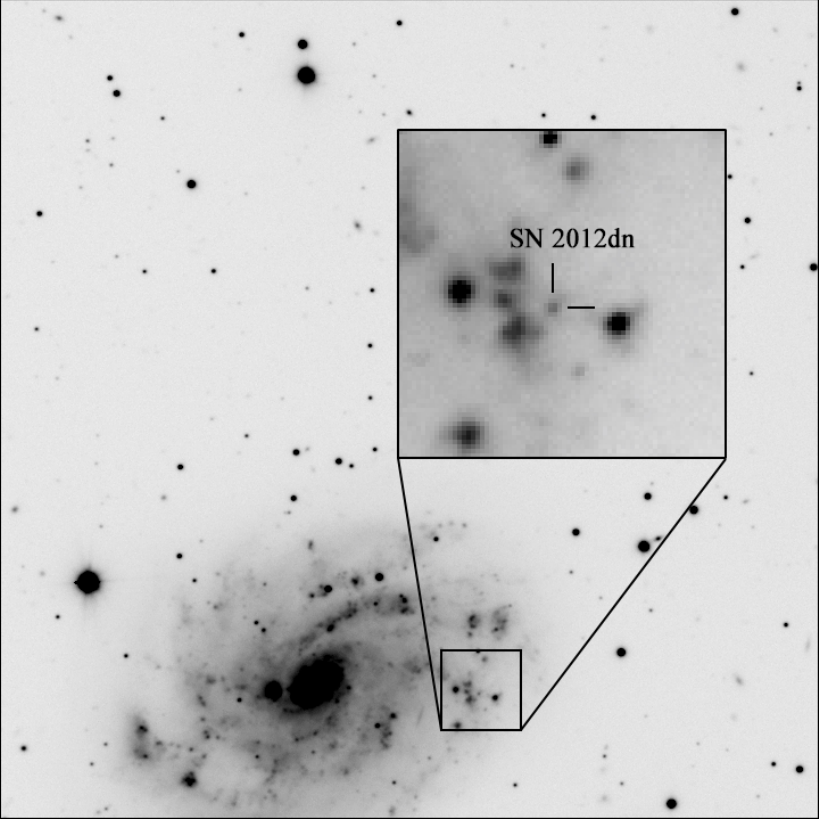}
  \caption{VLT\,+\,FORS2 $R$-band image of SN~2012dn, obtained 288\,d after maximum light. The field of view is $3 \times 3$ arcmin$^2$, north is up, east to the left, and the region around the SN is enlarged in an inset. }
  \label{fig:image}
\end{figure}

About a year after the explosion, the SNfactory follow-up of SN~2012dn was complemented by observations at the ESO 8.2m Very Large Telescope (VLT-UT1 and -UT3) on Cerro Paranal, Chile, equipped with the focal-reducer spectrograph and camera FORS2 and the Infrared Spectrometer And Array Camera ISAAC. Both imaging data and spectroscopy were obtained (Fig.~\ref{fig:image}).

The $BV\!RI$-band frames were debiased and flatfield-corrected, the $JH$-band frames sky-subtracted and combined, all using standard \textsc{iraf}\footnote{\textsc{iraf} is distributed by the National Optical Astronomy Observatory, which is operated by the Association of Universities for Research in Astronomy under cooperative agreement with the National Science Foundation.} tasks. The SN magnitudes were measured with \textsc{snoopy}\footnote{\textsc{snoopy} is a collection of scripts, originally developed by F. Patat and later implemented in \textsc{iraf} by E. Cappellaro. It is optimised for PSF-fitting photometry of point sources superimposed on a structured background.} using PSF photometry. Photometric errors were estimated performing an artificial-star experiment. For the $BV\!RI$ images, the instrumental zero points were obtained from a standard-field observation taken during the same night under photometric conditions, and the SN magnitudes were calibrated to the rest-frame Bessell \citep{bessell2012a} photometric system by means of $S$- and $K$-corrections \citep{nugent2002a,stritzinger2002a}. The $JH$-band frames were calibrated to the 2MASS photometric system \citep{cohen2003a} using eight 2MASS stars in the field. The final calibrated magnitudes and their errors, calculated as the quadratic sum of photometric and calibration errors, are reported in Tables~\ref{mags} and \ref{IRmags}.

Spectra were obtained on MJD = 56423.3 and MJD = 56448.4 (290.9 and 316.0\,d after $B$-band maximum) with grisms 300V and 300I (+\,OG590) and a 1\,arcsec slit, resulting in a wavelength coverage from $\sim$3400 to $\sim$10\,200\,\AA\ and a resolution of $\sim$9\,\AA. After low-level CCD corrections, the spectra were extracted using a variance-weighted algorithm within the \textsc{iraf} task \textsc{apall}. Wavelength calibration was accomplished with the help of arc-lamp exposures and if necessary adjusted slightly through a constant offset to match the position of night-sky emission lines. Observations of spectrophotometric standard stars were used to establish a relative flux calibration and remove telluric absorptions in the SN spectra. The absolute flux calibration of the spectra was finally checked against the SN photometry, but only small adjustments were necessary.

\section{Analysis of early-time data}
\label{Analysis of early-time data}

\subsection{Distance and extinction revisited}
\label{Distance and extinction revisited}

Previous studies have revealed significant variety in the photometric and spectroscopic properties of 09dc-like SNe~Ia, suggesting that it is not a good idea to use the colour evolution to determine the reddening \citep[e.g., table 4 in][]{scalzo2012a}. We therefore concentrate on interstellar absorption lines to estimate the extinction along the line of sight. Narrow interstellar Na\,D  lines are visible in all our spectra, and we measure equivalent widths (EWs) of $0.40\pm0.04$\,\AA\ and $0.33\pm0.06$\,\AA\ in a co-added spectrum ($-$13.2 to +15.7\,d) for the Galactic and host-galaxy components, respectively. Using the relation of \citet{turatto2003a}, this translates into $E(B-V)_\mathrm{MW} = 0.064\pm0.006$ and \linebreak[4] $E(B-V)_\mathrm{host} = 0.053\pm0.010$ mag for the colour excesses in the Milky Way and the host galaxy, respectively. The relation by \citet{poznanski2012a} yields values of $E(B-V)_\mathrm{MW} = 0.041\pm0.008$ and $E(B-V)_\mathrm{host} = 0.035\pm0.008$ instead. The Milky Way values are in fair agreement with the Galactic extinction measurement towards ESO\,462-016 by \citet{schlafly2011a} [$E(B-V) = 0.054$ mag]. We hence adopt a host-galaxy colour excess of $E(B-V)_\mathrm{host} = 0.044\pm0.013$ (the average of the Turatto et al. and Poznanski et al. based estimates), and a total reddening towards SN~2012dn of $E(B-V)_\mathrm{tot} = 0.10\pm0.02$ mag. This is 0.08 mag smaller than the estimate by \citet{chakradhari2014a}, who measured larger EWs for both the Galactic and host-galaxy Na\,D lines in one of their spectra, but slightly larger than the upper limit on the reddening of 0.09 mag estimated by \citet{parrent2016a} based on their Na\,D EW measurement and the \citet{poznanski2012a} relation. Assuming $R_V = 3.1$, our adopted colour excess results in a moderate total $V$-band extinction along the line of sight of $0.31\pm0.06$ mag.

The distance of ESO\,462-016 is not constrained by Cepheid measurements. A recent Tully-Fisher measurement by \citet{lagattuta2013a} yielded a distance of $43.7 \pm 13.5$ Mpc ($\mu = 33.19 \pm 0.69$ mag). This is in good agreement with kinematic distances \citep{mould2000a} of $43.9 \pm 4.2$ Mpc ($\mu = 33.21 \pm 0.21$ mag; corrected for infall onto the Virgo cluster) or $47.5 \pm 4.2$ Mpc ($\mu = 33.38 \pm 0.20$ mag; corrected for infall onto the Virgo cluster, the Great Attractor and the Shapley supercluster). All these numbers have been converted assuming $H_0 = 70$ km\,s$^{-1}$\,Mpc$^{-1}$, and the uncertainties in the kinematic distances have been computed assuming a peculiar velocity up to 300 km\,s$^{-1}$ for ESO\,462-016. For the rest of this paper, we use an average distance modulus $\mu = 33.26 \pm 0.20$ mag as our best estimate for SN~2012dn, in agreement with the 33.15 mag adopted by \citet{chakradhari2014a} and \citet{parrent2016a}, and the 33.32 mag adopted by \citet{brown2014a}.

\begin{figure*}
  \centering 
  \includegraphics[width=17.6cm]{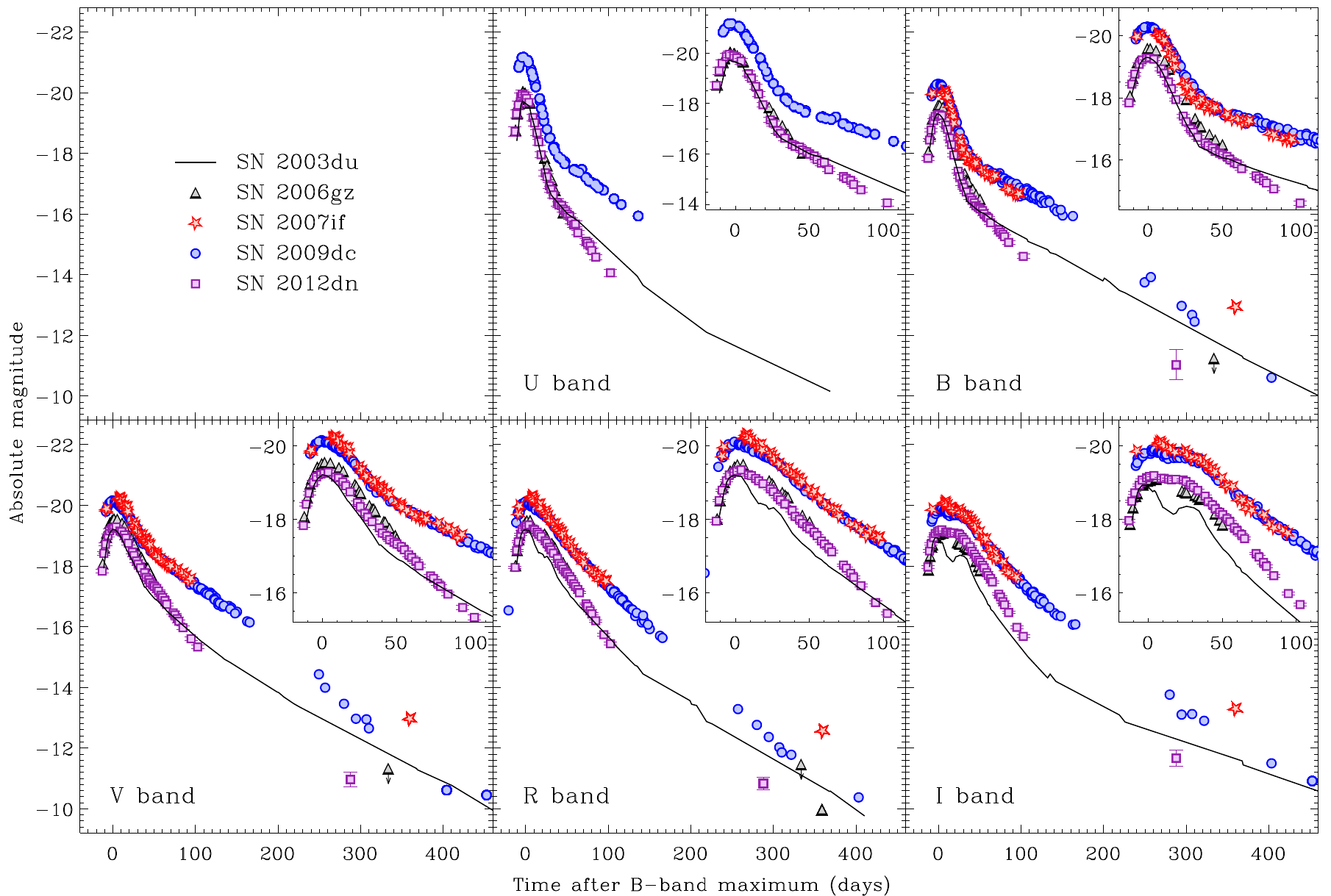}
  \caption{Absolute-magnitude $U\!BV\!RI$ light curves of SN~2012dn, the 09dc-like SNe~2006gz \citep{hicken2007a,maeda2009a}, 2007if \citep{scalzo2010a,taubenberger2013a} and 2009dc \citep{silverman2011a,taubenberger2011a}, and the normal SN~Ia 2003du \citep{stanishev2007b}. Error bars are only shown for SN~2012dn and do not include the uncertainties in distance and extinction. Downward-pointing arrows denote upper limits.}
  \label{fig:phot}
\end{figure*}

\begin{figure*}
  \centering 
  \includegraphics[width=17.6cm]{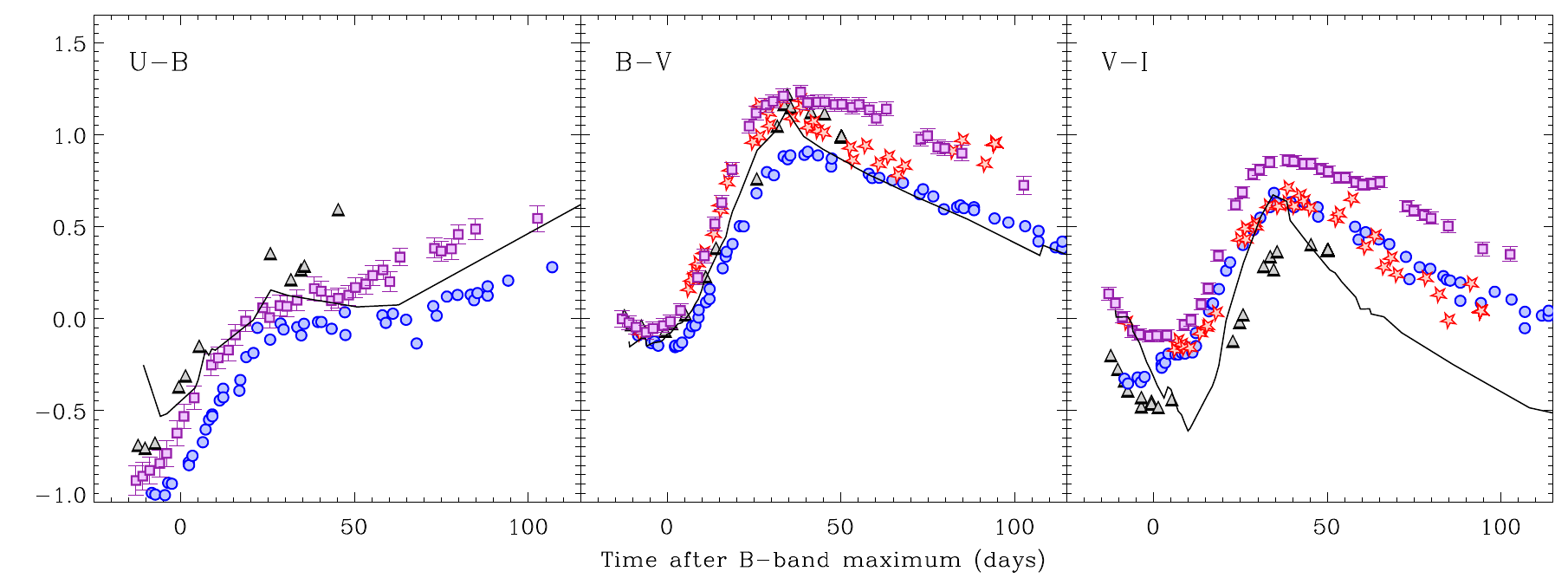}
  \caption{$U-B$, $B-V$ and $V-I$ colour evolution of the same SNe as in Fig.~\ref{fig:phot}. Error bars are only shown for SN~2012dn and include only photometric measurement errors but no uncertainties in extinction.}
  \label{fig:colours}
\end{figure*}

\begin{figure}
  \centering 
  \includegraphics[width=8.4cm]{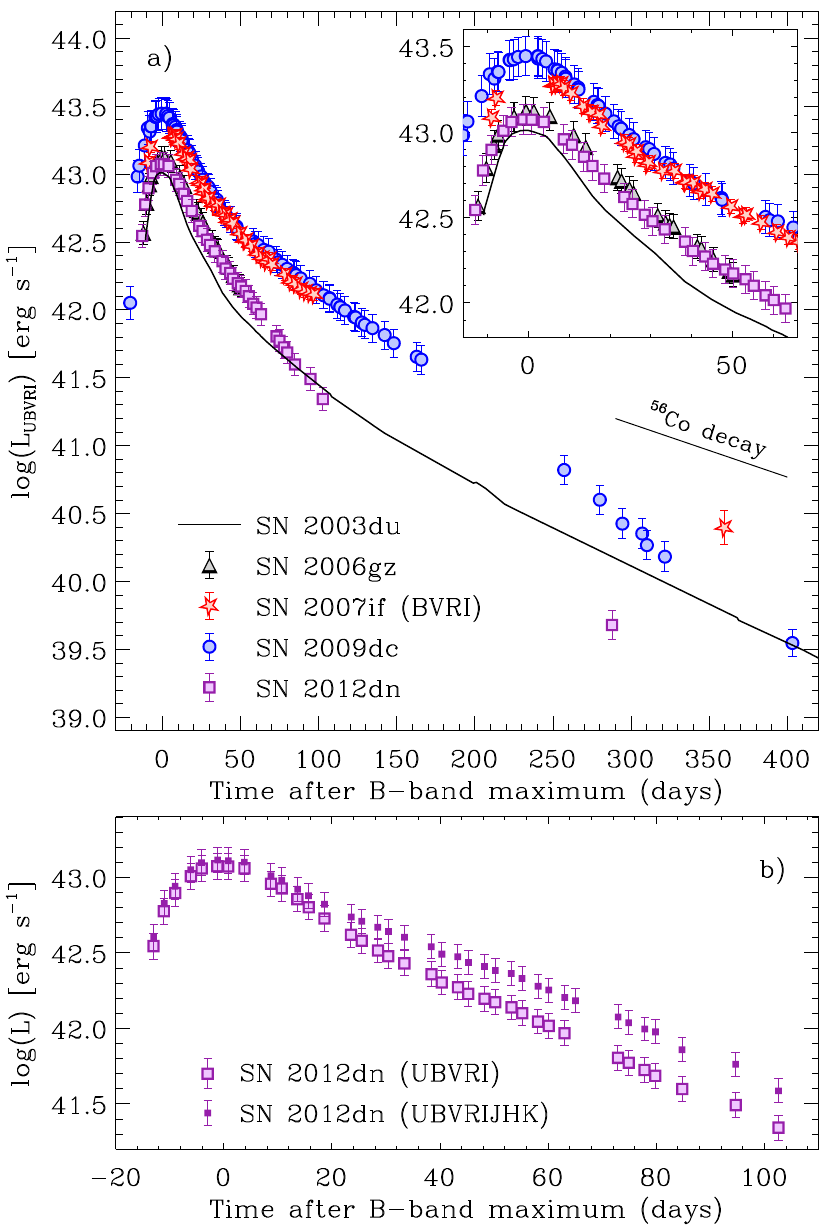}
  \caption{a) $U$-through-$I$ pseudo-bolometric light curves of SNe~2012dn, 2006gz, 2007if, 2009dc and the normal SN~Ia 2003du. Uncertainties in distance and extinction dominate the error budget in all these objects. The inset in the upper right corner contains an enlargement of the early phase up to 60\,d post-peak. b) Comparison of the $U$-through-$I$ and $U$-through-$K$ pseudo-bolometric light curves of SN~2012dn. The latter have been computed using our optical data combined with the NIR data presented by \citet{yamanaka2016a}.}
  \label{fig:bolo}
\end{figure}

\begin{figure}
  \centering 
  \includegraphics[width=8.4cm]{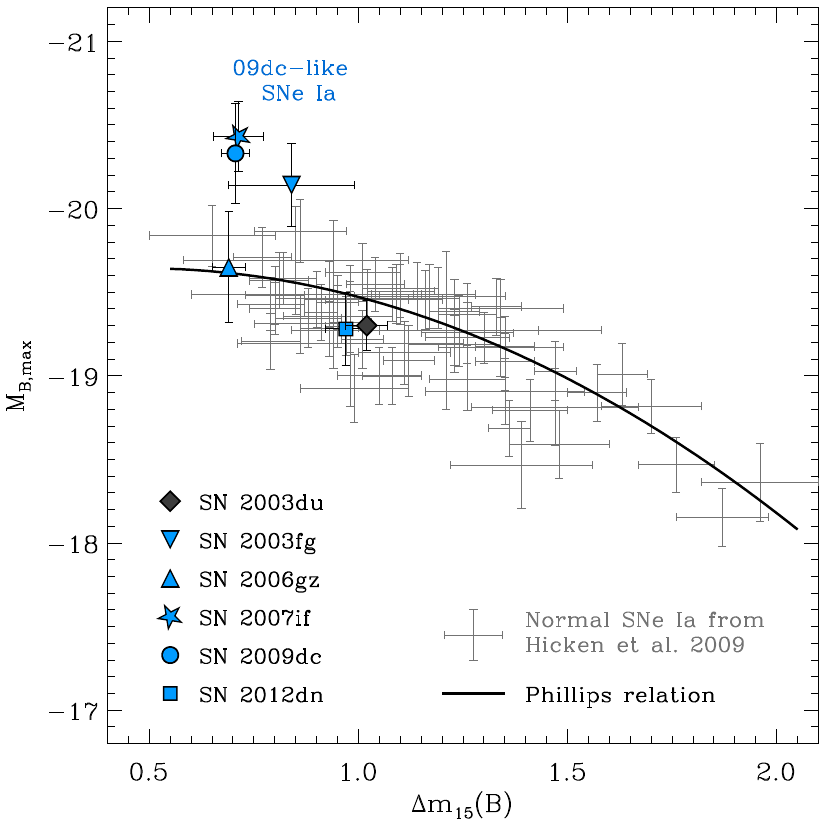}
  \caption{The location of 09dc-like SNe~Ia relative to the Phillips relation \citep{phillips1999a} and a subset of the CfA3 sample of normal SNe~Ia \citep{hicken2009b} and the normal SN~Ia 2003du \citep{stanishev2007b}.}
  \label{fig:Phillips}
\end{figure}

\subsection{Light curves and colour evolution}
\label{Light curves and colour evolution}

Fig.~\ref{fig:phot} shows the photometric evolution of SN~2012dn, together with those of the 09dc-like SNe~2006gz \citep{hicken2007a,maeda2009a}, 2007if \citep{scalzo2010a,taubenberger2013a} and 2009dc \citep{silverman2011a,taubenberger2011a}, and the normal SN~Ia 2003du \citep{stanishev2007b}. Fig.~\ref{fig:colours} compares the $U-B$, $B-V$ and $V-I$ colour evolution of the same five SNe, and Fig.~\ref{fig:bolo} shows how their $U$-through-$I$-band pseudo-bolometric luminosity evolves. For the comparison SNe, the pseudo-bolometric flux was calculated by translating the reddening-corrected filter magnitudes into monochromatic fluxes at the filters' effective wavelengths, interpolating linearly and integrating the resulting spectral energy distribution (SED) from the blue cut-on of the $U$-band filter to the red cut-off of the $I$-band filter. For SN~2012dn, we directly integrated the flux-calibrated, reddening-corrected spectra over the corresponding wavelength range (3300--8800\,\AA; Scalzo et al. 2014). From the comparison in Figs.~\ref{fig:phot} to \ref{fig:bolo}, a number of properties become evident, several of them already discussed by \citet{chakradhari2014a}, \citet{brown2014a} and \citet{parrent2016a}:
\begin{enumerate}

\item The morphology of the SN~2012dn filter light curves resembles more closely that of other 09dc-like SNe~Ia than those of normal or 91T-like SNe~Ia. This is particularly evident in the red and NIR bands, where SN~2012dn -- like SNe~2006gz, 2007if and 2009dc -- does not show two distinct maxima, but a single broad peak. The $I$-band maximum does not precede that in $B$ as in normal SNe~Ia.

\item The light-curve peak of SN~2012dn (MJD$_{\mathrm{max},B} = 56132.6 \pm 0.2$) is somewhat narrower and declines more rapidly than in other 09dc-like SNe~Ia. We measure a reddening-corrected \dm15$_\mathrm{true}$ \citep{phillips1999a} of 0.97, compared to 0.69, 0.71 and 0.71 for SNe~2006gz, 2007if and 2009dc, respectively. This is not far from that of normal SNe~Ia such as SN~2003du (1.02).

\item Like other 09dc-like SNe~Ia and unlike normal SNe~Ia, SN~2012dn has a very blue pre-maximum $U-B$ colour, a trend which is also seen in other UV\,$-$\,optical colours \citep{brown2014a}. In other colour indices such as $B-V$ or $V-I$, and especially at later epochs, SN~2012dn appears slightly redder than other 09dc-like SNe~Ia. 

\item With the distance and extinction estimated in Section~\ref{Distance and extinction revisited}, SN~2012dn reaches a $B$-band peak absolute magnitude of $-19.28\pm 0.22$ and a $U$-through-$I$ pseudo-bolometric peak luminosity of $1.19 \times 10^{43}$ erg\,s$^{-1}$, very similar to normal SNe~Ia, and fainter than SN~1991T or other 09dc-like SNe~Ia. While SN~2012dn still appeared mildly overluminous with the extinction estimated by \citet{chakradhari2014a}, this is no longer the case with the revised reddening estimates used by \citet{parrent2016a} and us. In fact, while all other 09dc-like SNe~Ia lie `above' the Phillips relation \citep{phillips1999a} [in the sense of being more luminous than predicted by their \dm15], SN~2012dn lies slightly below the nominal Phillips line, at a very similar locus as SN~2003du (Fig.~\ref{fig:Phillips}).

\item About 60\,d after maximum, soon after the light curves of SN~2012dn seem to have settled onto a radioactive tail, they start to fade more rapidly again. This behaviour was already noticed by \citet{chakradhari2014a}, and is now confirmed by our more extended photometry. However, the effect is subtle, and can only be appreciated by comparing the light-curve evolution during the tail phase to those of other SNe, e.g. to the normal SN~Ia 2003du. We will further discuss this behaviour on the basis of late-time observations in Section~\ref{Late-time light curve}.

\end{enumerate}

\subsection{Spectroscopic evolution}
\label{Spectroscopic evolution}

\begin{figure*}
  \centering 
  \includegraphics[width=12.0cm]{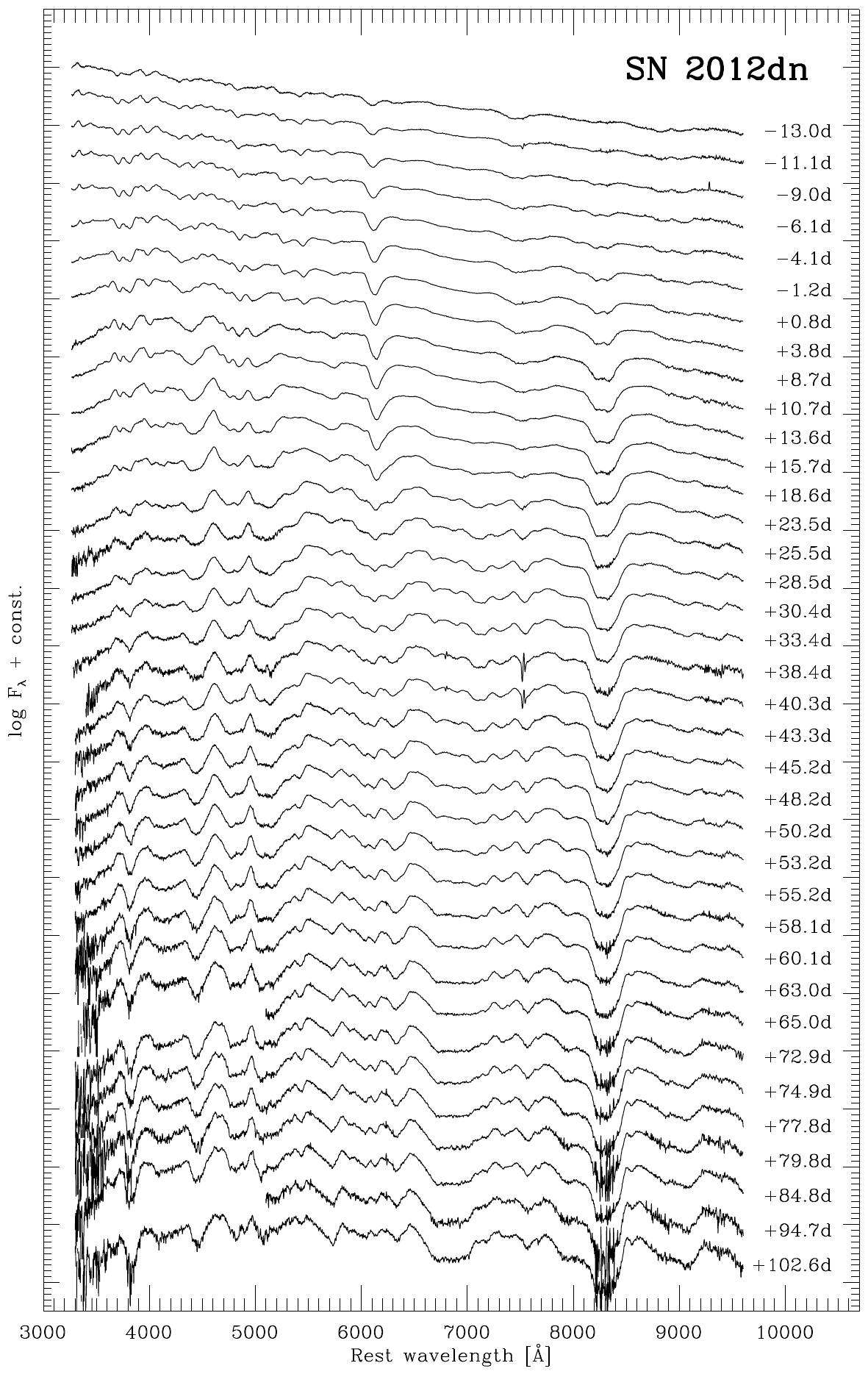}
  \caption{Time series of SNIFS spectra of SN~2012dn, from very early phases to $\sim$100\,d after $B$-band maximum (MJD = $56132.6\pm0.2$).}
  \label{fig:spectra}
\end{figure*}

\begin{figure*}
  \centering 
  \includegraphics[width=14.5cm]{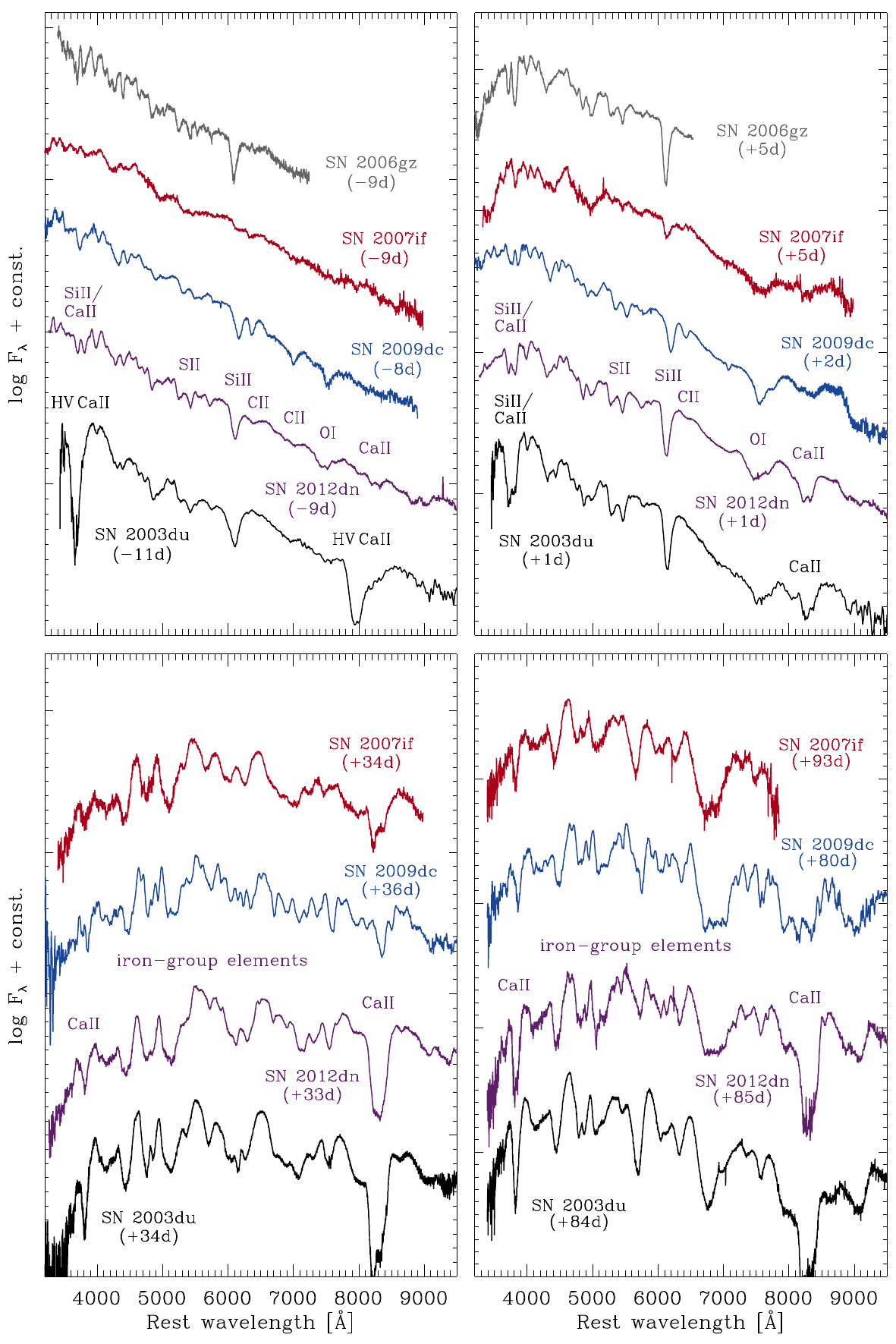}
  \caption{Comparison of the spectra of SN~2012dn with those of the 09dc-like SNe 2006gz \citep{hicken2007a}, 2007if \citep{scalzo2010a} and 2009dc \citep{taubenberger2011a}, and the normal SN~Ia 2003du \citep{stanishev2007b} at 10\,d before $B$-band maximum, around maximum, a month after maximum, and three months after maximum, respectively. All spectra have been corrected for interstellar extinction.}
  \label{fig:spectra_comp}
\end{figure*}

Our full time series of SN~2012dn spectra obtained with SNIFS, ranging from $-13.0$ to $+102.6$\,d with respect to $B$-band maximum light, is presented in Fig.~\ref{fig:spectra}. This data set constitutes the most extensive homogeneous spectroscopic time series of any 09dc-like SN to date. In Fig.~\ref{fig:spectra_comp} the spectra of SN~2012dn at selected epochs are compared with those of the 09dc-like SNe~2006gz, 2007if and 2009dc, and the normal SN~Ia 2003du. 

The spectra of SN~2012dn at pre-maximum phases are different from those of normal SNe~Ia such as SN~2003du in several respects. They show a high UV flux, especially at the earliest epochs where the peak of the SED lies outside the observed wavelength range. This finding is consistent with the high early-time UV flux inferred from \textit{Swift} photometry by \citet{brown2014a}. \CII\ lines ($\lambda6580$ and $\lambda7234$) are clearly detected, and persist beyond maximum light \citep{chakradhari2014a,parrent2016a}. All other spectral features except for prominent \OI\ $\lambda7774$ are relatively weak and narrow in SN~2012dn. \SiII\ $\lambda6355$ and \CaII\ lines are particularly weak, and show no high-velocity components. 
The described characteristics are very typical of 09dc-like SNe~Ia, and the early spectra of SN~2012dn closely resemble those of SNe~2006gz and 2009dc. However, while the line strengths and widths are similar in these three objects, there is diversity in the line velocities calculated from the blueshift of P-Cygni absorptions. SNe~2006gz and 2012dn show much higher line velocities than SN~2009dc, as already shown by \citet{chakradhari2014a} and discussed further in Section~\ref{Velocity evolution}.

After maximum light, the spectra of SN~2012dn become increasingly similar to those of normal SNe~Ia with normally broad spectral features due to iron-group elements and a prominent \CaII\ NIR triplet. Those of SN~2009dc, in contrast, remain peculiar, and with their narrow lines they resemble those of SNe~Iax \citep[][and references therein]{foley2013a}.

\subsection{Velocity evolution of the line-forming region}
\label{Velocity evolution}

\begin{figure}
  \centering 
  \includegraphics[width=8.4cm]{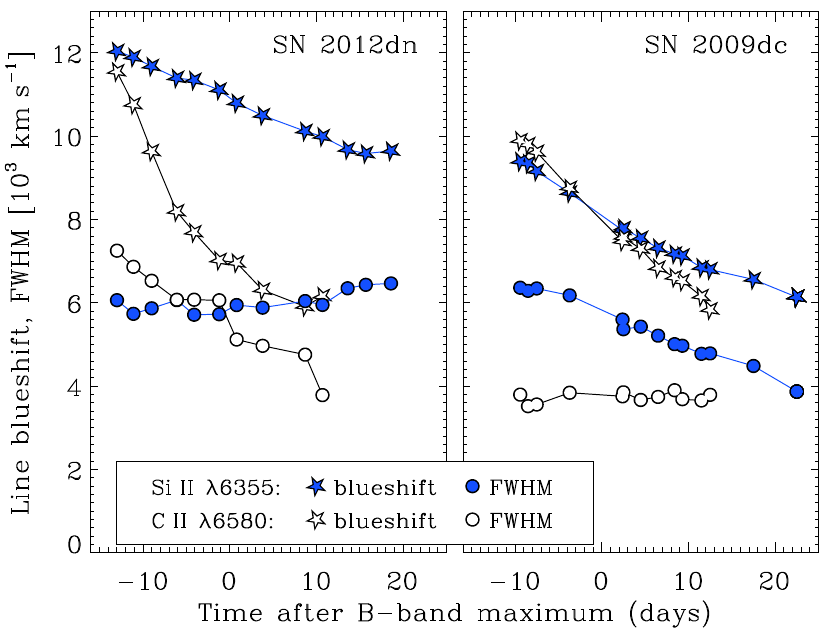}
  \caption{Time evolution of blueshifts and widths of the \SiII\ $\lambda6355$ and \CII\ $\lambda6580$ lines in SNe~2012dn and 2009dc, respectively.}
  \label{fig:velocities}
\end{figure}

To examine more closely the location of the line-forming region in SNe~2012dn and 2009dc, Fig.~\ref{fig:velocities} shows the blueshift and full width at half-maximum (FWHM) of the \SiII\ $\lambda6355$ and \CII\ $\lambda6580$ absorption lines as a function of time. Both quantities were estimated from Gaussian fits to the absorption components of the respective P-Cygni lines, and carry independent information: the line position is an indicator for the velocity coordinate of strongest line formation, while the line width probes the extent of the line-forming region in velocity space. 
In most SNe, there is a close correlation between line blueshift and FWHM, in the sense that strongly blueshifted absorptions are also wide, but there are notable exceptions, and SNe~2012dn and 2009dc are two of them.

The blueshift of \CII\ and \SiII\ decreases with time in both SNe, interpreted as a line-forming region that recedes deeper into the ejecta as a consequence of the expansion. However, while in SN~2009dc the blueshifts of \SiII\ and \CII\ are similar at all times, in SN~2012dn they are only similar at very early phases, but then the \CII\ blueshift decreases much more rapidly than that of \SiII. This is particularly interesting, since explosion models would generically predict an unburned element like carbon to be located in layers exterior to those rich in intermediate-mass elements (IMEs) such as silicon. Nevertheless, already by the time of maximum light, the \CII\ line velocity in SN~2012dn is 4000\,\kms\ lower than that of \SiII, and in both SNe \CII\ absorption is detected down to a velocity of $\sim6000$\,\kms.

The picture becomes even more complicated when considering the FWHM evolution of the lines. For \CII\ in SN~2012dn and \SiII\ in SN~2009dc, the FWHM of the lines decreases with time, approximately in parallel to the decreasing blueshift (that the FWHM is generally smaller than the blueshift is caused by the emission component of P-Cygni lines, and hence not unexpected). However, \SiII\ in SN~2012dn and \CII\ in SN~2009dc show a completely flat FWHM evolution with time, despite their decreasing blueshift. In these lines, the difference between blueshift and FWHM is particularly large, especially at early phases. Hence, we are facing a narrowly confined line-forming region at a relatively large radius, i.e. `detached' lines in a simple photospheric picture. In their SYNAPPS \citep{thomas2011a} modelling of SN~2012dn spectra, \citet{parrent2016a} also found that two weeks before maximum essentially all spectral lines in SN~2012dn are detached, with typical detachment velocities of 12\,000--14\,000\,\kms. If the inner boundaries of the line-forming regions were  dictated by the elemental abundance profiles of IMEs and unburned material, the lines could not recede further into the ejecta at later times (which they do). Hence, they are probably, at least in part, determined by the ionisation state.

At pre-maximum epochs, the \SiII\ lines in SNe~2012dn and 2009dc have the same FWHM of $\sim6000$\,\kms, but the blueshift of the line in SN~2012dn is 2500--3000\,\kms\ larger than in SN~2009dc. Conversely, the \CII\ lines in the two objects have similar blueshift (8000--10\,000\,\kms), but very different FWHM (6000--7000\,\kms\ in SN~2012dn vs. 4000\,\kms\ in SN~2009dc). All in all these results reveal considerable diversity in the abundance and\,/\,or ionisation structure of the ejecta of 09dc-like SNe~Ia.

\section{Analysis of late-time data}
\label{Analysis of late-time data}

\subsection{Late-time light curve: rapid decline and low luminosity}
\label{Late-time light curve}

As pointed out by \citet{chakradhari2014a} and mentioned in Section~\ref{Light curves and colour evolution}, the light curves of SN~2012dn show an enhanced decline, compared to normal SNe~Ia or SN~2009dc, starting around 60\,d after maximum. An additional late-time photometric data point obtained with VLT\,+\,FORS confirms this trend: at 287\,d after maximum SN~2012dn is between 0.7 ($I$ band) and 1.5\,mag ($B$ and $V$ bands) fainter than SN~2003du (Fig.~\ref{fig:phot}). We note that photometry between 330 and 350\,d past maximum reported by \citet{parrent2016a} suggests SN~2012dn to be about 2 mag brighter than in our FORS images. This is perplexing, as it seems unlikely that such a dramatic rebrightening can occur on a time scale of 1--2 months, especially since the spectroscopic evolution between +290 and +315\,d (Section~\ref{Nebular spectrum}) is smooth and inconspicuous. We can therefore only speculate that the photometry of \citet{parrent2016a} might be heavily contaminated by host-galaxy light (SN~2012dn exploded in a complex region of its host with a spiral arm and strong \HII-region emission in its vicinity) or that the SN was confused with a different object. 

Relying on our late-time photometry, the average decline rate between 100 and 287\,d past maximum ranges from $\sim$1.9\,mag\,(100\,d)$^{-1}$ in the $B$ band to $\sim$2.5\,mag\,(100\,d)$^{-1}$ in the $R$ band [$\sim$2.2\,mag\,(100\,d)$^{-1}$ in the pseudo-bolometric light curve], which is not only much steeper than \Cofs\ decay [0.98\,mag\,(100\,d)$^{-1}$], but also significantly steeper than the decline of SN~2003du in the same bands at similar phases, which ranges from 1.0 and 1.7\,mag\,(100\,d)$^{-1}$ \citep{stanishev2007b}.
% B: 1.93 mag/100d  V: 2.36 mag/100d  R: 2.49 mag/100d  I: 2.17 mag/100d  bol: 2.19 mag/100d

\citet{chakradhari2014a} speculated about newly formed dust as the origin of the enhanced fading, motivated by dust formation claimed for SNe~2006gz and 2009dc \citep{maeda2009a,taubenberger2013a}. However, while in SN~2009dc the rapid fading sets in about 200\,d after maximum (and in SN~2006gz the lack of light-curve coverage does not allow us to determine the instant precisely), an onset only two months after maximum in SN~2012dn is remarkably early. Possible explanations for the enhanced decline will be discussed in Section~\ref{Late optical light curve}.

\subsection{Nebular spectrum: low ionisation, emission at 6300\,\AA}
\label{Nebular spectrum}

\begin{figure}
  \centering 
  \includegraphics[width=8.4cm]{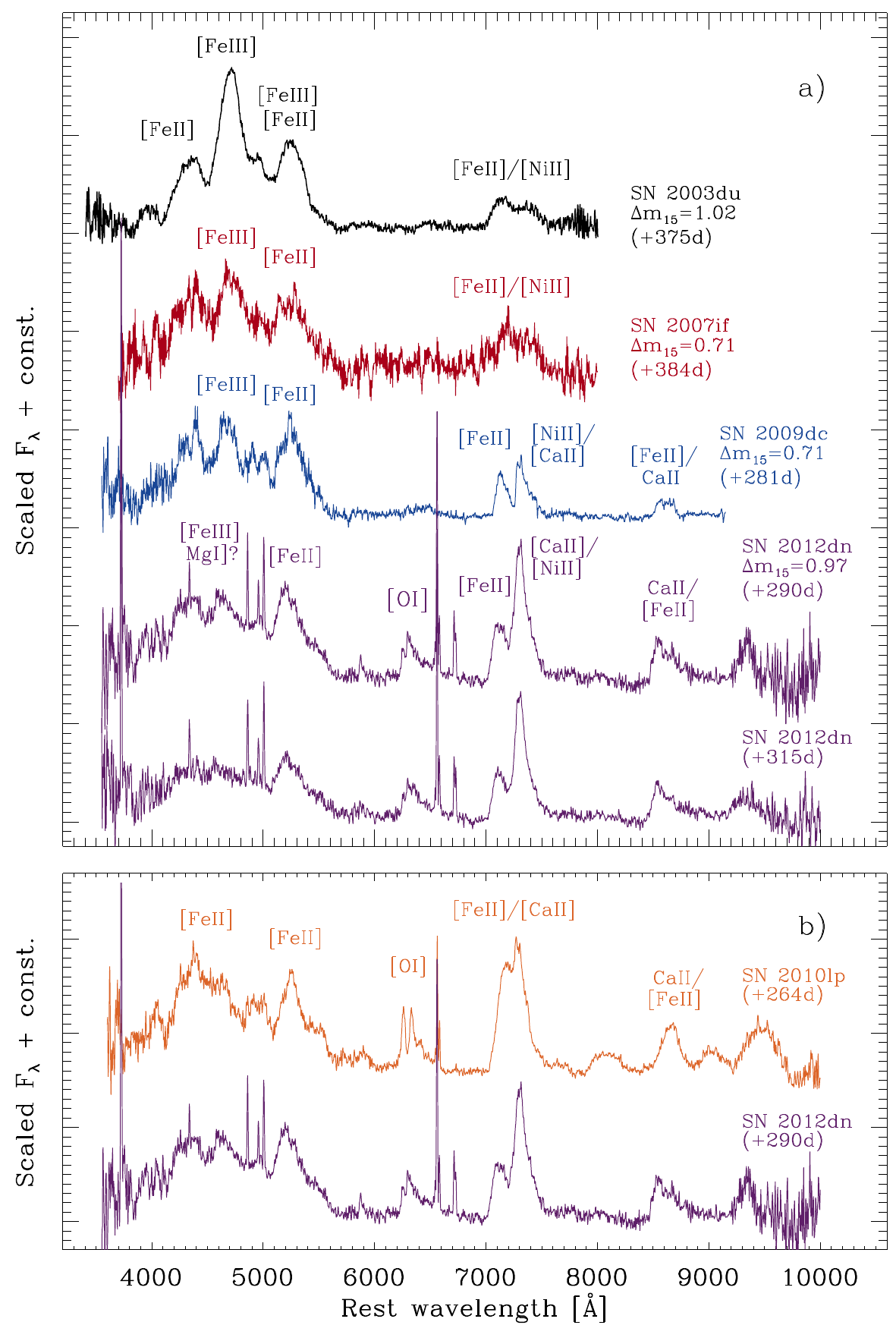}
  \caption{a) Nebular spectra of SN~2012dn compared to those of the 09dc-like SNe~2007if \citep{taubenberger2013a} and 2009dc \citep{silverman2011a}, and the normal SN~Ia 2003du \citep{stanishev2007b}. b) A comparison with SN~2010lp \citep{taubenberger2013b} supports the identification of [\OI] $\lambda\lambda6300,6364$ emission in the nebular spectrum of SN~2012dn. All spectra have been corrected for interstellar extinction. Strong, narrow \HII-region emission lines are apparent in the SN~2012dn spectra.}
  \label{fig:nebular}
\end{figure}

\begin{figure}
  \centering 
  \includegraphics[width=8.4cm]{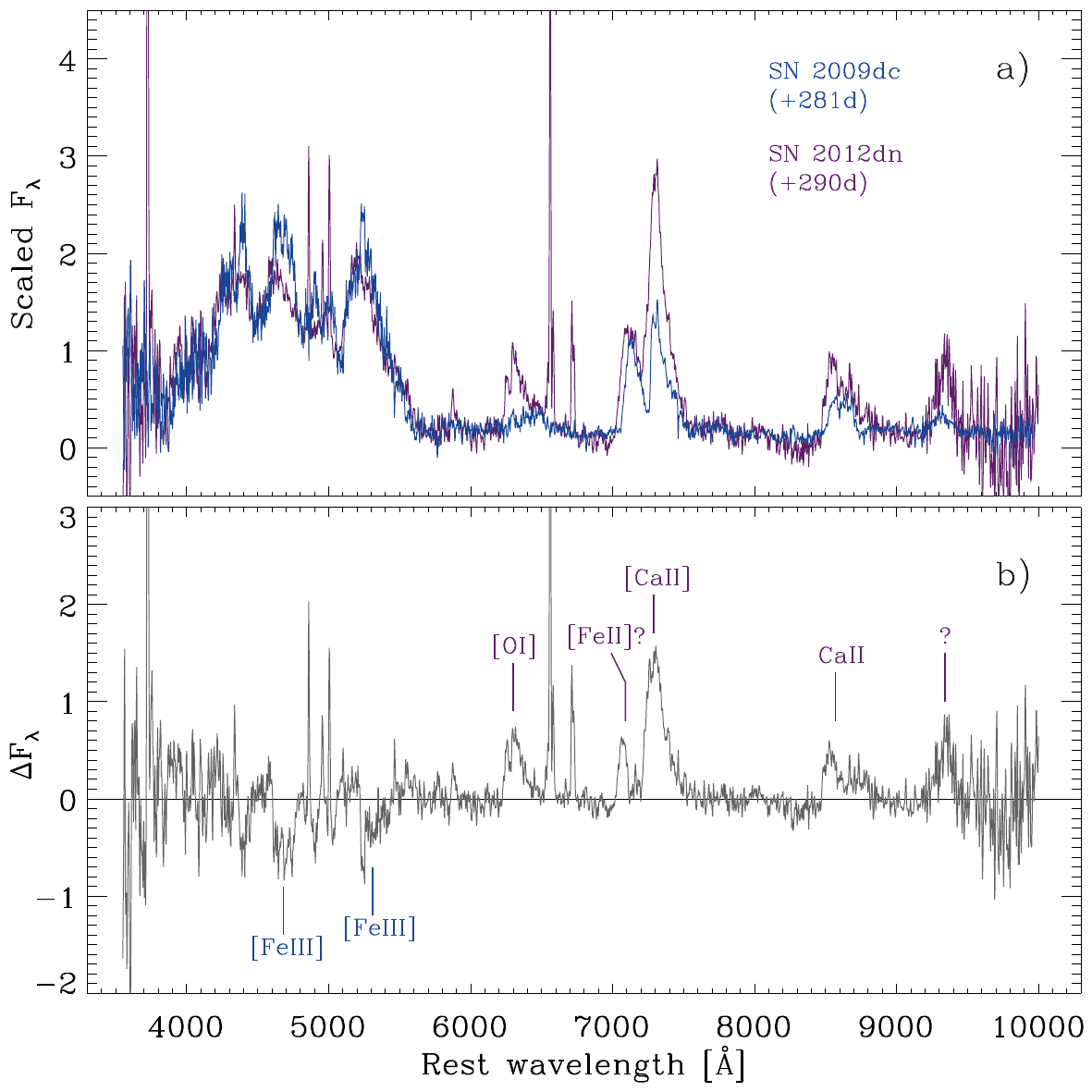}
  \caption{a) Flux-matched nebular spectra of SNe~2012dn and 2009dc. The overall similarity of the spectra is strong, though a few distinct differences are evident. b) The subtraction of the scaled spectra from the upper panel highlights the differences between SNe~2012dn and 2009dc: weaker [\FeIII], but stronger calcium and oxygen emission in SN~2012dn.}
  \label{fig:diff_12dn_09dc}
\end{figure}

Nebular spectra of SN~2012dn obtained 290 and 315\,d after $B$-band maximum are shown in Fig.~\ref{fig:nebular}a and compared to similarly late spectra of the normal SNe~Ia 2003du \citep{stanishev2007b} and the 09dc-like SNe~Ia 2007if and 2009dc \citep{taubenberger2013a}. As in other 09dc-like SNe~Ia, [\FeIII] emission around 4700\,\AA\ is weak in SN~2012dn, whereas [\FeII] is clearly detected all over the spectrum, suggesting low ejecta temperatures or a high ejecta density which favours recombination and hence a low ionisation state. In fact, since the recent nebular spectrum-synthesis calculations by \citet{botyanszki2017a} suggest the dominant ionisation state of iron to be \FeIV\ in normal SNe~Ia around 200\,d, the difference in the ionisation state of 09dc-like SNe~Ia has to be quite dramatic in order to suppress [\FeIII] emission.

While [\FeII] dominates the nebular spectra of SN~2012dn, there are other emission lines that are absent in nebular spectra of other SNe~Ia -- including other 09dc-like events (cf. Fig.~\ref{fig:diff_12dn_09dc}) -- or at least not nearly as prominent. This includes a strong feature at $\sim$7300\,\AA, which could be interpreted as the [\NiII] $\lambda7378$ line. However, for that to be true it would have to be strongly blueshifted by $\gtrsim$3000\,\kms, 1500--2000\,\kms\ more strongly than [\FeII] lines. For this reason, we are confident that the feature is formed by [\CaII] $\lambda\lambda7292,7324$ instead, with a FWHM of $\sim$4000\,\kms, and a blueshift of only 300\,\kms. Even more strikingly, there is a prominent emission feature around 6300\,\AA, reminiscent of those seen in the subluminous SNe~Ia 2010lp and iPTF14atg and interpreted as the [\OI] $\lambda\lambda6300,6364$ line by \citet{taubenberger2013b} and \citet{kromer2013b,kromer2016a}, respectively (Fig.~\ref{fig:nebular}b). If the [\OI] identification in SN~2012dn is correct, the line has a FWHM of $\sim$5500\,\kms\ (similar to that estimated for the [\FeII] lines) and a negligible red- or blueshift. 

In SN~2010lp and iPTF14atg the identification with [\OI] is plausible, given the low luminosity of these objects and the accordingly low burning efficiency, although the oxygen would still have to be rather centrally located in order to explain the observed line profiles.
The identification is more challenging to reconcile with expectations for an 09dc-like SN~Ia, even for a moderately luminous one such as SN~2012dn. Another SN~Ia for which emission near 6300\,\AA\ has been detected is SN~2011fe, but there the feature was only present in very late spectra after $\sim$1000\,d \citep{graham2015a,taubenberger2015a}. In that case, the discussion revolved around whether this was [\OI] $\lambda\lambda6300,6364$ or instead \FeI\ emission \citep{taubenberger2015a}. The successful modelling as a non-thermal \FeI-dominated spectrum by \citet{fransson2015a} solved this controversy in favour of \FeI. The nebular spectra of SN~2012dn, however, deviate from the 1000\,d spectra of SN~2011fe and the \citet{fransson2015a} models in showing clear evidence for emission from thermally excited lines, e.g. the low-lying, forbidden 7155\,\AA\ line of \FeII, which is not part of typical recombination cascades. \FeI\ apparently does not dominate the spectra overall, and it is unclear from atomic physics why exactly and exclusively the \FeI\ lines around 6300\,\AA\ should emerge in SN~2012dn, leaving [\OI] $\lambda\lambda6300,6364$ as the most plausible identification for the 6300\,\AA\ line.

Interpreting line profiles in nebular spectra is generally complicated by line blending, but judging from a few weakly blended lines in the red part of the spectrum, the profiles in SN~2012dn appear sawtooth-shaped, with a steeper blue edge and a more gentle slope on the red side. Such a profile could arise from non-spherical ejecta with an emissivity distribution significantly peaked towards the near side of the ejecta, possibly caused by an off-centre \Nifs\ distribution. Such a geometry could also be responsible for some of the observed peculiarities in the early-time velocity evolution of SN~2012dn (see Section~\ref{Spectroscopic evolution}). Alternatively, similar line profiles have been observed in other SNe after the formation of dust in the ejecta \citep{lucy1989a}, when the emission from the receding far side of the SN is more strongly extinguished than that coming from the near side. This second possibility could lend support to speculations about dust formation in SN~2012dn based on the light curve evolution, which will be discussed more thoroughly in Section~\ref{Discussion}.

\subsection{\HII-region metallicity estimate}
\label{Metallicity}

Previous studies \citep{khan2011a} have suggested that 09dc-like SNe~Ia prefer metal-poor environments. To test this for SN~2012dn, we use the strong \HII-region emission lines superimposed on the nebular spectra (Fig.~\ref{fig:nebular}a), which allow us to estimate the gas-phase metallicity at the location of SN~2012dn. We re-extracted the spectra, applying no local background subtraction, but rather choosing the background outside the host-galaxy to subtract only the telluric night-sky emission. We set the extraction aperture to 1\,arcsec diameter, corresponding to 217\,pc at the assumed distance of ESO~462-016. The reddening towards the \HII\ region was estimated from the Balmer decrement assuming case-B recombination. The reddening derived in this way amounts to $E(B-V) = 0.16 \pm 0.02$\,mag, consistent with some estimates for SN~2012dn found in the literature \citep{chakradhari2014a}, although there is no reason that the two line-of-sight extinctions should match exactly. Flux ratios of various lines were then measured in reddening-corrected spectra.

The derived [\OIII]/H$\beta$, [\NII]/H$\alpha$ and [\SII]/H$\alpha$ ratios place the environment of SN~2012dn in the \HII-region portion of BPT \citep*{baldwin1981a} diagrams, with no evidence for shocks \citep[e.g.][]{kewley2006a}. Following \citet{curti2017a}, we use six different indicators to determine the local gas-phase metallicity through strong-line diagnostics. Some of them, like R23 or O32, are strongly dependent on the adopted extinction, since they involve ratios with the [\OII] $\lambda3727$ line in the near UV, but for the moderate extinction in our case the effect is still quite small. Other indicators, such as R3, N2 and O3N2, involve only flux ratios of lines with similar wavelength and are thus rather insensitive to the extinction. All indicators show good agreement and suggest the metallicity at the site of SN~2012dn to be $12+\log(\mathrm{O/H}) = 8.57 \pm 0.05$, corresponding to $Z = 0.51\pm0.06\,Z_\odot$ if a solar oxygen abundance of $12+\log(\mathrm{O/H})_\odot = 8.86$ \citep{delahaye2010a} is adopted. This moderately low metallicity is in agreement with the findings that 09dc-like SNe~Ia prefer metal-poor environments, although it is not nearly as low as for the extreme case of SN~2007if with $12+\log(\mathrm{O/H}) = 8.01 \pm 0.09$ or $Z = 0.15\,Z_\odot$ \citep{childress2011a}. It remains uncertain, however, to what extent the present-day gas-phase metallicity is representative for the metallicity of the SN~2012dn progenitor system. Even though there are indications that stellar orbits exhibit strong correlations for a Gyr or more \citep[see the discussion in][]{rigault2013a,rigault2015a}, if the progenitor system of SN~2012dn was long-lived, the gas in its environment may have been chemically enriched by several generations of core-collapse SNe in the meantime. This leads to a potential metallicity difference between the SN~2012dn progenitor and nearby \HII\ regions, thus turning our estimate into an upper limit on the SN metallicity.

\section{Discussion}
\label{Discussion}

\subsection{Late optical light curve: $\gamma$-ray and positron escape, IR catastrophe or dust formation?}
\label{Late optical light curve}

Only a few mechanisms can lead to an accelerated optical light-curve decline: $\gamma$-ray and positron escape \citep{ruiz-lapuente1998a}, the termination of interaction with a circumstellar or circumbinary medium, an IR catastrophe (IRC; \citealt{axelrod1980a,fransson2015a}) and dust formation in the ejecta. Among those, $\gamma$-ray and positron escape and the discontinuation of interaction processes are true bolometric effects, whereas an IRC or dust formation merely lead to a redistribution of flux out of the observed optical bands towards longer wavelengths.\smallskip

\textit{$\gamma$-ray and positron escape:} 
For a radioactively driven light curve and under the assumption of full positron trapping and complete $\gamma$-ray leakage, the pseudo-bolometric luminosity of SN~2012dn measured at +287\,d, $\log(L_{U\!BV\!RI}) = 39.68$, can be converted to a \Nifs\ mass of $\sim$\,0.16\,\msun. Even allowing for a NIR contribution to the bolometric luminosity of 50 per cent (hence doubling the \Nifs\ mass estimate to 0.32\,\msun), this number is significantly smaller than the \Nifs\ mass of $\sim$\,0.8\,\msun\ one would derive from the light-curve peak applying Arnett's rule \citep{arnett1982a}. To remedy this discrepancy, an unusually early leakage of the positrons from \Cofs\ decay might be invoked, with a trapping fraction of merely 40 per cent at +287\,d.
The trapping of positrons depends on the strength and orientation of magnetic fields in the ejecta \citep{ruiz-lapuente1998a}. Tangled magnetic fields lead to very efficient trapping for long times, and in fact even after $\geq$\,1000\,d no evidence for positron escape has been found in normal SNe~Ia \citep{kerzendorf2014a,kerzendorf2017a,graur2016a,shappee2017a,dimitriadis2017a}. But even in the case of totally absent or perfectly radial magnetic fields, the trapping time scale for positrons is about 20 times longer than that for $\gamma$-rays \citep{jerkstrand2017a}. Using the formula provided by \citet{jerkstrand2017a} for the positron-trapping time scale in a magnetic-field free, homogeneous sphere, $t_{\mathrm{trapping,}e^+} = 620\,\mathrm{d}\ (M\,/\,1\,\msun)\ (E\,/\,10^{51}\,\mathrm{erg})^{-1/2}$, and setting $E/M\,\approx\,0.8$ (in units of $10^{51}$\,erg\,$/\,\msun$) in line with the measured ejecta velocities, the total ejecta mass could not be larger than $\sim$\,0.23\,\msun\ to comply with the maximum allowed 40 per cent positron trapping at +287\,d estimated above. This is a clear contradiction, since already the \Nifs\ mass derived from the light-curve peak is three to four times larger.

A possible solution could be that the assumption of a homogeneous sphere is strongly violated. In particular, strong clumping in the ejecta could reduce the mean optical depth substantially for a given ejecta mass. However, the decay $\gamma$-rays and optical photons would feel the same reduction in optical depth as the positrons. Hence, to achieve the maximum allowed 40 per cent positron trapping at +287\,d for a \Nifs\ mass of 0.8\,\msun, the ejecta would have to turn optically thin to $\gamma$-rays already $\sim$2\,weeks after the explosion. A broad peak and a bright early light-curve tail as observed in SN~2012dn would not be feasible under these conditions.

The only way to reconcile the clumping scenario with the light curve of SN~2012dn is to assume that clumping developed only after a certain point in time. At that moment, the mean optical depth and hence the $\gamma$-ray and positron deposition would drop, and an accelerated light-curve decline would be observed. However, in the fast ballistic flow of SN ejecta, clumping cannot develop spontaneously. The only conceivable way to generate clumping as late as 60\,d after maximum would be the interaction with a patchy circum- or interstellar medium. This, however, should leave clear fingerprints in the spectra and produce additional light for the duration of the interaction, both of which are not observed in SN~2012dn.\smallskip

\textit{Discontinued circumstellar or circumbinary interaction:} In this case the first 2--3 months of the light curve of SN~2012dn would not be exclusively driven by the decay of \Nifs, but by ongoing conversion of kinetic energy to light through interaction processes. If this additional power source switched off around 60\,d past maximum, the bolometric luminosity of SN~2012dn would drop rapidly, converging eventually to the radioactive tail dictated by the \Nifs\ decay chain. The luminosity measured at +287\,d would be the intrinsic luminosity of the SN, and the derived \Nifs\ mass of 0.16\,M$_\odot$ would be an estimate of the true Ni mass produced by SN~2012dn. IR corrections to the late-time pseudo-bolometric luminosity would increase the \Nifs-mass estimate, residual $\gamma$-ray trapping would decrease it. In any case, the \Nifs\ mass would be very low and make SN~2012dn an intrinsically strongly subluminous event, meaning that the vast majority of the light during the peak and early tail phase would be contributed by circumstellar\,/\,circumbinary interaction. While this scenario of an intrinsically subluminous SN whose light curve is boosted by interaction processes during the first 2--3 months is hard to exclude rigorously, it is not favoured by our observations, especially by the post-maximum spectroscopic evolution that bears resemblance with normal SNe~Ia (Fig.~\ref{fig:spectra_comp}).\smallskip

\textit{Infrared catastrophe:} The IRC is a process that usually takes place in SN ejecta at late times. It requires a temperature below 2000\,K, and during the IRC the temperature rapidly drops further to only a few hundred K. This is accompanied by far-IR fine-structure transitions taking over most of the cooling, and by a significant change in the plasma state and the optical spectrum, as evidenced in SN~2011fe between 300 and 1000\,d after explosion \citep{graham2015a,taubenberger2015a,fransson2015a}. In SN~2012dn no unexpected evolution of the optical spectrum is observed between +60 and +100\,d compared to normal or other 09dc-like SNe~Ia (Fig.~\ref{fig:spectra_comp}), and even the nebular spectra at $\sim$+300\,d are still dominated by strong [\FeII] emission e.g. at 7155\,\AA, consistent with thermal excitation in a plasma of a few thousand K (Fig.~\ref{fig:nebular}). This is inconsistent with an IRC having occurred in the Fe core. In principle, the same thermal instability should also take place in the IME-rich layers of the SN, but since they contribute only very little to the total emission at $t \geq +60$\,d, a change there cannot be responsible for an enhanced fading of the light curves. Moreover, the likely detection of [\OI] $\lambda\lambda6300,6364$ in the nebular spectra of SN~2012dn indicates that also the IME-rich layers are still in the thermal regime after 300\,d, since [\OI] $\lambda\lambda6300,6364$ is not part of an \OI\ recombination cascade \citep{jerkstrand2015a}. This is corroborated by the strength of [\CaII] $\lambda\lambda7292,7324$: in the limit of pure non-thermal excitation, the ratio of [\CaII] $\lambda\lambda7292,7324$ to the \CaII\ NIR triplet should approach unity \citep{jerkstrand2017a}, which is clearly not the case in SN~2012dn at 300\,d.\smallskip

\textit{Dust formation:} The formation of graphite or silicates is quite common in core-collapse SNe after a few hundred days, and has also been suggested to have occurred in the 09dc-like SNe~2006gz and 2009dc to explain the low optical luminosities measured about one year after the explosion \citep{maeda2009a,taubenberger2013a}. In particular, \citet{taubenberger2013a} suggested that the dust in SN~2009dc formed in a carbon-rich, dense shell, located outside the Fe core where the SN spectrum forms at those late phases. The optical luminosity of SN~2012dn at a comparable epoch is similarly or even more suppressed (Fig.~\ref{fig:bolo}) than in SNe~2006gz and 2009dc, but in contrast to SN~2009dc where deviations from the normal \Cofs\ tail start only after +200\,d, in SN~2012dn the fading gradually starts around day +60. Such an early onset of dust formation would be unprecedented even among normal core-collapse SNe (where the conditions for dust formation are supposedly much more favourable than in thermonuclear SNe owing to the lower amount of radioactive material), and is observed only in a few peculiar SNe such as the interacting SN~Ibn 2006jc \citep{pastorello2007c,mattila2008b,smith2008a}. In SN~2006jc, the inferred site of dust formation was a cool dense shell, formed by the interaction of the SN ejecta with a dense CSM \citep{mattila2008b,smith2008a}. The dust in SN~2006jc was likely comprised of graphite grains \citep{smith2008a}, and dust condensation started about 50\,d after explosion \citep{mattila2008b}, signalled by the onset of a flux excess in the NIR bands, and followed by a rapid fading of the optical light curves with $\sim$20\,d of delay. This behaviour resembles that of SN~2012dn in almost every detail. Even the lack of conspicuous colour evolution in the optical regime after the onset of dust formation (cf. \citealt{pastorello2007c} and Fig.~\ref{fig:colours}) is a common theme in both SNe and can plausibly be explained by a combination of absorption and scattering processes on the embedded dust grains, or a very clumpy dust distribution.

\subsection{IR excess: pre-existing or newly formed dust?}
\label{IR excess}

\citet{yamanaka2016a} presented NIR light curves of SN~2012dn, which show strong deviations from those of normal SNe~Ia or SN~2009dc from $\sim$\,30\,d after maximum light onwards. Instead of a smooth and relatively steep decline as in the latter, the $J$ and $H$-band light curves of SN~2012dn show a delayed decline, and in the $K_\mathrm{S}$ band even a significant rebrightening (see fig.~3 of \citealt{yamanaka2016a}). The subtraction of templates constructed from the $JHK_\mathrm{S}$ light curves of SN~2009dc revealed an excess in the NIR bands which starts to rise at $\sim$\,30\,d after $B$-band maximum, reaches a peak after 70\,d, and declines very slowly thereafter (\citealt{yamanaka2016a}, fig.~8). A blackbody fit to the SED of the NIR excess yielded a temperature decreasing from 2200\,K at the onset of the excess to 1200\,K after 100\,d.

\citet{yamanaka2016a} and \citet{nagao2017a,nagao2018a} interpreted this excess as a NIR echo caused by pre-existing circumstellar dust around SN~2012dn. The dust grains would have to be made of amorphous carbon, since silicates would evaporate at the given temperatures. \citet{nagao2017a} considered two possible dust distributions (a disk- and a bipolar-jet-like geometry), and from the onset of the NIR excess they inferred an inner cavity of 0.04\,pc radius, roughly consistent with the evaporation radius for graphite dust as a consequence of the SN radiation field at maximum light. 

Yamanaka et al. and Nagao et al. also discussed the possibility that the NIR excess could be caused by newly formed dust in the ejecta of SN~2012dn. However, they dismiss this scenario based on two arguments: the short time scales, with an onset about 50\,d after the explosion, and the fact that the optical light curves show no unexpected colour evolution at those phases. While these are certainly valid concerns, the argument neglects the fact that all optical light curves show an enhanced decline (Fig.~\ref{fig:phot}) roughly coincident in time with the onset of the NIR excess, while the echo models of \citet{nagao2017a} instead predict a slight flattening at that phase due to a weak optical light-echo contribution (see their fig.~5). The fading is also visible in the full $U$-through-$K$-band bolometric light curve (Fig.~\ref{fig:bolo}b), and may reflect the progressive shift of the thermal emission out of the observed bands as the temperature decreases below 2000\,K. We would rather argue along the following lines:
\begin{enumerate}
\item As discussed in Section~\ref{Late optical light curve}, the observed dimming of the optical light curves by itself might be most readily explained by an unusually early formation of dust in the ejecta.

\item A thermal NIR excess which develops at a similar time would be a natural consequence of the dust-formation scenario, so we take this as supporting evidence. In fact, even that the onset of the emission slightly precedes the measurable extinction might possibly be understood in this scenario, since the first dust grains form at a high temperature of $\gtrsim2000$\,K and thus emit strongly, while their extinction is still negligible.

\item One could in principle imagine a situation where a combination of dust formation and an unrelated echo by pre-existing dust is encountered, but following Occam's razor a single scenario that explains both the optical dimming and the NIR excess appears more compelling.

\item New dust formation and subsequent cooling would naturally explain the observed temperature evolution of the thermal IR emission \citep{yamanaka2016a}, which is less straightforward in an echo model where different physical regions may contribute to the emission at different times.

\end{enumerate}

To explore the scenario of dust formation and the possible location and geometry of the dust in a more quantitative way, we make some estimates based on calculations performed by \citet{yamanaka2016a} and \citet{nagao2017a}. In particular, we want to determine the radius $R$ of a hypothetical dust shell from the measured extinction and thermal emission by assuming radiative equilibrium. We start from fig.~10 of \citet{yamanaka2016a}, where the dust mass and temperature  have been derived from a blackbody fit to the SED of the $H$- and $K$-band flux excess of SN~2012dn. At +80\,d, the dust emits at a temperature of 1300\,K, and the estimated dust mass is $\sim$10$^{-4}$\,\msun\ for a dust-grain radius $a = 0.01$\,$\mu$m.
If this dust is located in a thin spherical shell, the optical depth can be calculated as in \citet{taubenberger2013a}:
\begin{equation}
\label{eq:tau}
\tau_\nu = \frac{Q_\nu(a) \pi a^2 M_\mathrm{dust}(a)}{4 \pi R^2 m_\mathrm{grain}}
\end{equation}
Amorphous carbon has a density of 2.2\,g\,cm$^{-3}$, and for a dust-grain radius of 0.01\,$\mu$m the extinction efficiency in the $V$ band, $Q_V$, is $\sim$0.07 \citep{draine1984a}. With an estimated $V$-band extinction of $\sim$0.4\,mag at +80\,d from newly formed dust (corresponding to $\tau_V=0.37$), we can solve Eq.~\ref{eq:tau} to derive a dust-shell radius $R = 0.010$\,pc. Assuming a SN rise time of 20\,d (so that a phase of 80\,d after $B$-band maximum corresponds to a time of 100\,d from explosion), this radius can be translated to a velocity in homologous expansion of 37\,000\,\kms.

This radius is very similar to the evaporation radius for amorphous carbon of $\sim$0.007\,pc at the time of the beginning of dust formation (50\,d after $B$-band maximum). The latter can be estimated by starting from the evaporation radius of 0.02\,pc derived by \citet{nagao2017a} for the peak of the SN~2012dn radiation field and rescaling it to account for the lower luminosity and redder SED at +50\,d (using our pseudo-bolometric light curve and the absorptive opacities of table~1 of \citealt{nagao2017a}). At +50\,d, 0.007\,pc correspond to a velocity in homologous expansion of 36\,000\,\kms.

Both estimates depend on the assumed dust-grain size, but only weakly. Varying the average dust-grain size within reasonable limits, the derived radii would not change by more than 20--30 per cent. In any case, these locations lie well outside the ejecta, and any dust there can not be due to the SN. To reconcile the idea of new dust formation and the observational constraints with a location of the dust in a shell at much lower velocity, one needs a way for the dust to survive the intense radiation field and to result in a lower extinction for a given dust mass. Both could be achieved if the dust was not distributed in a homogeneous thin shell, but in dense clumps. In this case, self-shielding of the optically thick clumps would protect a large fraction of the dust grains from the SN radiation, and the succession of regions with very large optical depth and regions with zero optical depth would result in an overall smaller attenuation of the SN light. As a by-product, such a configuration with quasi geometrical blocking of part of the SN light would lead to a nearly grey extinction over wide wavelength ranges, and hence be a natural explanation for the observed lack of chromatic extinction in SN~2012dn in the optical bands. Moreover, contrary to a thin dust shell whose optical depth would decrease as $t^{-2}$ as the shell expands, the dense dust clumps would remain optically thick, and the extinction would be time-independent after dust formation has ceased, which appears more consistent with the late-time photometry of SN~2012dn.

As a final discriminant between pre-existing and newly formed dust, one can resort to the prediction of \citet{nagao2017a} that in the case of pre-existing dust an additional contribution to the optical emission should be observable at late phases, arising from the interaction of the SN ejecta with the surrounding circumstellar medium (CSM). With our VLT photometry after 300\,d, we can test this prediction to some extent. If SN~2012dn did not explode in a true CSM cavity, but only the dust within the inner 0.04\,pc was evaporated by the SN radiation, there should be a certain level of ejecta--CSM interaction at all phases, which should dominate the optical luminosity after some point in time. Precisely, Nagao et al. predict an $R$-band magnitude of about 21.0 after 400\,d or about 22.5 after 500\,d, depending on which of their preferred CSM geometries applies. With our measured $R$-band magnitude of 22.7 at about 300\,d after the explosion, both models can be rejected. Instead, a true inner cavity of the CSM would be required. If the CSM starts only at 0.04\,pc, the interaction would only set in about 1000\,d after the explosion and at a much fainter level. Our photometry has no constraining power for this scenario.

\subsection{Implications for explosion scenarios}
\label{Implications for explosion scenarios}

A suitable explosion scenario should ideally provide a consistent explanation for all observed peculiarities, but at the same time be flexible enough to account for the diversity encountered among 09dc-like SNe~Ia. We argue that the observed diversity lends further support to the scenario proposed by \citet{hachinger2012a} and \citet{taubenberger2013a}: an enshrouded explosion inside an H- and He-deficient envelope. This envelope could be the leftover of a WD--WD merger, and would be swept up by the SN ejecta on time scales of hours to few days. In this model, the exploding WD itself would not exceed \MCh, and it should therefore not be regarded as a super-Chandrasekhar scenario\footnote{The explosion dynamics, kinetic energy and nucleosynthesis yields are very different from those of the explosion of a genuine super-Chandrasekhar WD as studied by \citet{pfannes2010b,pfannes2010a} and \citet{fink2018a}.}, even though the total ejecta mass including the swept-up envelope may eventually exceed \MCh. By varying the properties of the underlying explosion, the mass of the envelope, its density profile, and the viewing angle in case of a non-spherical envelope configuration, substantial diversity can be realised. At the same time, some generic properties such as a boost in bolometric luminosity, a deceleration of the SN ejecta and broadening of the light curves by the reverse shock, an enhancement in unburned C and O abundances, and a perturbed asymptotic density structure with a dense C- and O-rich shell as a possible place of dust formation are retained. 

\citet{noebauer2016a} investigated this scenario more carefully, running radiation-hydrodynamical models to study the effect of a brief, early interaction episode on the SN light curve, the SED and ejecta density profile. They were able to reproduce several characteristics of 09dc-like SNe~Ia quite well, but failed to reproduce the boost in optical luminosity required to match SN~2009dc owing to an insufficient redistribution efficiency of the higher-energy radiation produced by the shock (which might, however, be related to shortcomings in the atomic data and opacity treatment in the simulation). 

In contrast, scenarios in which the peak luminosity is exclusively provided by the decay of a fairly large amount of \Nifs\ are once more disfavoured by late-time data of SN~2012dn \citep[cf.][for the same discussion in the context of SN~2009dc]{taubenberger2013a}. The low ionisation state of the ejecta inferred from late-time spectra, with [\FeIII] lines being essentially absent, requires low temperatures and high densities to favour recombination. At the same time, lines from IMEs and even oxygen are likely detected, which are extremely weak or absent in normal SNe~Ia. This also points at a low temperature of the nebula, and at an unusually high abundance of lighter elements. Moreover, dust formation as early as two months after the explosion would not be possible if the \Nifs\ mass was very high and the heating by $\gamma$-rays very strong.

\subsection{Defining characteristics of the class of 09dc-like SNe Ia}
\label{Defining characteristics of the class of 09dc-like SNe Ia}

\begin{figure}
  \centering 
  \includegraphics[width=8.4cm]{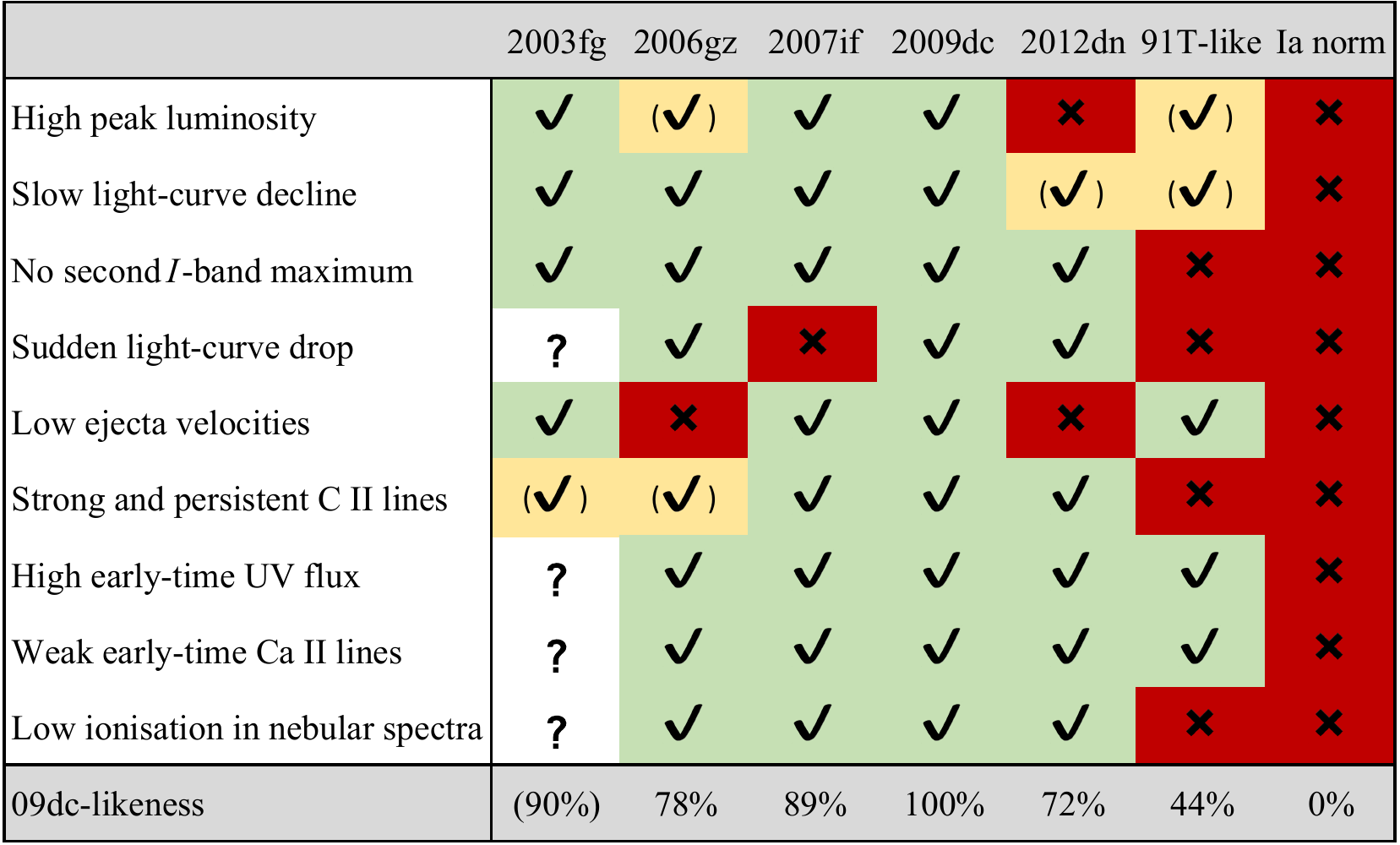}
  \caption{Criteria for membership in the class of 09dc-like objects (see Section~\ref{Defining characteristics of the class of 09dc-like SNe Ia}) applied to SNe~2003fg, 2006gz, 2007if, 2009dc, 2012dn, 91T-like SNe and normal SNe~Ia. Green stands for `09dc-like', red for `not 09dc-like', and yellow for `transitional'. Within this scheme, objects denoted as 09dc-like SNe in this paper are `09dc-like' in 70 to 100 per cent of all categories (with `transitional' evaluations getting half credit).    
  Note that for SN~2003fg only five out of nine criteria were used, since the limited data do not allow an assessment of the other categories. The result for SN~2003fg is thus not directly comparable to the other SNe.}
  \label{fig:09dc_likeness}
\end{figure}

From the photometric and spectroscopic data it is evident that SN~2012dn is in several respects less extreme than other 09dc-like SNe~Ia. In particular, with the revised host-galaxy reddening estimate discussed in Section~\ref{Distance and extinction revisited} it is not any more luminous than many slowly declining normal SNe~Ia. From an unbiased analysis of the peak light curves and spectra there is no evidence to support an extraordinarily high \Nifs\ or ejecta mass. Still, compared to normal SNe~Ia, SN~2012dn has numerous peculiarities that clearly connect it to more extreme 09dc-like SNe~Ia such as SN~2009dc itself \citep{yamanaka2009a,silverman2011a,taubenberger2011a}. Throughout this work, we have therefore refrained from describing the class of objects as `super-Chandrasekhar SNe~Ia', which refers to a possibly incorrect physical interpretation, or `superluminous SNe~Ia', which emphasises a frequent but not universal property, and instead chosen to use the more neutral term `09dc-like SNe~Ia'. However, the definition of a separate class only makes sense if common characteristics can be identified. For 09dc-like SNe, it appears that no single criterion does justice to the variety encountered among the candidate class members, and that only the use of a whole set of criteria provides a sufficiently informed view on potential 09dc-likeness. The criteria we intend to use have been proposed by various authors in earlier works. They are briefly described below and applied in Fig.~\ref{fig:09dc_likeness}.
\begin{enumerate}
\item \textit{High peak luminosity.} Most 09dc-like SNe~Ia show a high peak luminosity. We define $M_{B,\mathrm{peak}} < -19.9$ as 09dc-like; $-19.9 \leq M_{B,\mathrm{peak}} < -19.5$ as transitional; $M_{B,\mathrm{peak}} \geq -19.5$ as not 09dc-like.
\item \textit{Slow light-curve decline.} A broad light-curve peak is another criterion for class membership. \dm15 $\leq 0.8$: 09dc-like; $0.8 <$ \dm15 $\leq 1.0$: transitional; \dm15 $> 1.0$: not 09dc-like.
\item \textit{No second $I$-band maximum.} Contrary to normal and 91T-like SNe~Ia, 09dc-like objects do not show two distinct maxima in the $I$-band and most NIR bands. Instead, they have broad, plateau-like light curves in those filters. Single broad $I$-band peak: 09dc-like; two distinct $I$-band maxima: not 09dc-like.
\item \textit{Sudden light-curve drop.} Several 09dc-like SNe exhibit an unexpected, rapid fading of their optical light curves after some point in time during the radioactive-tail phase, possibly related to dust formation. Accelerated optical light-curve decline during the tail phase: 09dc-like; normal decline: not 09dc-like.
\item \textit{Low ejecta velocities.} 09dc-like SNe typically have relatively narrow lines and low ejecta velocities. Taking the velocity derived from the blueshift of the \SiII\ $\lambda6355$ line at maximum light as a benchmark, we set a cut at 10\,000\,\kms, corresponding approximately to the 10th percentile of the \SiII-velocity distribution of SNe~Ia (Lin et al., submitted). We define $v_\mathrm{Si} \leq 10\,000$\,\kms: 09dc-like;  $v_\mathrm{Si} > 10\,000$\,\kms: not 09dc-like.
\item \textit{Strong and persistent \CII\ lines.} Normal SNe~Ia sometimes show weak \CII\ features at very early phases, but they have faded away by maximum light \citep{thomas2011b,folatelli2012a}. In 09dc-like SNe, they are significantly stronger at comparable epochs, and persist until 1--2 weeks past maximum. Strong \CII\ up to maximum light, and still visible a week after maximum: 09dc-like; weaker or less persistent \CII: transitional; \CII\ only before maximum light or no \CII: not 09dc-like.
\item \textit{High early-time UV flux.} 09dc-like SNe are characterised by a high UV flux at early phases \citep{brown2014a}. At 7--10\,d before maximum, their SED peaks below 3500\,\AA, and the \CaII\ H\&K absorption is weak. SED peak below 3500\,\AA\ at $>1$ week before maximum: 09dc-like; SED peak above 3500\,\AA: not 09dc-like.
\item \textit{Weak early-time \CaII\ lines.} 91T-like SNe and 09dc-like SNe have in common, that their \CaII\ H\&K and NIR triplet lines before maximum light are much weaker than in normal SNe~Ia. Weak pre-maximum \CaII: 09dc-like; strong \CaII: not 09dc-like.
\item \textit{Low ionisation in nebular spectra.} The few nebular spectra available for 09dc-like SNe are characterised by unusually weak [\FeIII] lines compared to normal SNe~Ia \citep[e.g.][]{maguire2018a} or SN~1991T \citep{spyromilio1992a}, suggesting a low ejecta ionisation state. Weak [\FeIII]\,/\,low ionisation in nebular spectrum: 09dc-like; [\FeIII] dominates: not 09dc-like.
\end{enumerate}
Following this scheme, objects that are categorised as `09dc-like' in most individual criteria would be called 09dc-like SNe~Ia, since good agreement in so many different areas without a common underlying physical mechanism appears unlikely. As can be seen from Fig.~\ref{fig:09dc_likeness}, all the objects included in this work as 09dc-like SNe show at least 70 per cent agreement with the above catalogue of criteria (giving half credit to `transitional' evaluations), while 91T-like and normal SNe~Ia remain below 50 per cent agreement.

It goes without saying that the choice of criteria in this catalogue is to some degree subjective, and that the outcome of the classification process depends on the criteria used. For example, \citet{scalzo2012a} considered early-time spectra with prominent \FeIII\ lines and a flat evolution of the \SiII\ $\lambda6355$ velocity around maximum light to be important characteristics of their set of `super-Chandrasekhar' SNe~Ia (which, as already mentioned, is not identical with the set studied here). Also, following \citet{khan2011a} a low-metallicity environment could be used as an additional criterion for 09dc-like SNe. However, we point out that the commonalities among 09dc-like SNe as defined by our criteria are sufficiently strong that the overall picture does not change with limited adjustments to the criteria catalogue.

\section{Conclusions}
\label{Conclusions}

In this work, we have presented optical and NIR data of SN~2012dn, including the most extensive spectroscopic time series of any 09dc-like SN to date, as well as late-time photometry and spectroscopy taken about 1\,yr after the explosion. Based on light-curve morphology and spectroscopic peculiarities, the SN can clearly be identified as a member of the class of 09dc-like SNe~Ia, but with its rather modest luminosity and comparatively high ejecta velocities it adds significant diversity to the class. It thus provides motivation for a revised classification scheme which no longer relies on the luminosity as the primary feature. 

We have presented evidence for an enhanced fading of the optical light curves starting about 60\,d after maximum light -- a point in time roughly coincident with the excess emission at IR wavelengths reported by \citet{yamanaka2016a}. We argue that both effects are likely related, and probably caused by dust formation in parts of the ejecta, maybe in a carbon-rich dense shell. Such a shell could be the relic of an early, brief episode of interaction of the SN ejecta with a carbon- and oxygen-rich envelope \citep{taubenberger2013a,noebauer2016a}. Different masses and extensions of these envelopes, leading to different degrees of deceleration of the SN ejecta, might then explain the observed diversity of 09dc-like SNe~Ia in terms of ejecta velocities and late-time behaviour. To test the dust-formation hypothesis further, 09dc-like SNe need to be more extensively monitored at long wavelengths. We predict that NIR spectra will show CO overtone emission in the $K$ band, witnessing the formation of molecules, which pave the way for dust formation thanks to their enhanced cooling capabilities. Moreover, mid-IR photometry in addition to NIR photometry could show the cooling of the newly formed dust and the shifting of its thermal emission to ever longer wavelengths. 

Nebular spectra of SN~2012dn taken $\sim$300\,d after maximum reveal a low ionisation state, with nearly absent [\FeIII] emission. Lines near 6300 and 7300\,\AA\ can likely be attributed to [\OI] $\lambda\lambda6300,6364$ and [\CaII] $\lambda\lambda7292,7324$, which makes SN~2012dn a rare case of a SN~Ia where IMEs and unburned material are found in a location sufficiently close to radioactive material to be heated efficiently at late phases.

\section*{Acknowledgments}

This work is based on observations collected with ESO telescopes on Cerro Paranal, Chile, under programme IDs 091.D-0600 and 290.D-5035, and with the University of Hawaii 2.2\,m telescope on Mauna Kea, Hawaii. We thank the ESO support astronomers, the technical staff of the University of Hawaii 2.2\,m telescope, and Dan Birchall for observing assistance. We recognize the significant cultural role of Mauna Kea within the indigenous Hawaiian community, and we appreciate the opportunity to conduct observations from this revered site.
ST would like to thank Takashi Nagao and Richard Scalzo for helpful discussions, and the referee for insightful and constructive comments, which helped to improve the paper substantially.

This work has been supported by the Transregional Collaborative Research Center TRR33 `The Dark Universe' of the German Research Foundation, by the Cluster of Excellence `Origin and Structure of the Universe' at Technical University Munich, and the ESO Studentship Programme. Support in France was provided by CNRS/IN2P3, CNRS/INSU, and PNC; LPNHE acknowledges support from LABEX ILP, supported by French state funds managed by the ANR within the Investissements d'Avenir programme under reference ANR-11-IDEX-0004-02. Support in China was provided from Tsinghua University 985 grant and NSFC grant No 11173017.
In the U.S., this work was supported in part by the Director, Office of Science, Office of High Energy Physics, of the U.S. Department of Energy under Contract No. DE-AC02-05CH11231, and by a grant from the Gordon \& Betty Moore Foundation. Some results were obtained using resources and support from the National Energy Research Scientific Computing Center, supported by the Director, Office of Science, Office of Advanced Scientific Computing Research, of the U.S. Department of Energy under Contract No. DE-AC02-05CH11231. This project has received funding from the European Research Council (ERC) under the European Union's Horizon 2020 research and innovation programme (grant agreement No 759194 - USNAC).

\footnotesize{
  \bibliographystyle{mn2e}

}

\label{lastpage}

\end{document}